\documentclass{article}

\usepackage{arxiv}

\usepackage[utf8]{inputenc} % allow utf-8 input
\usepackage[T1]{fontenc}    % use 8-bit T1 fonts
\usepackage{hyperref}       % hyperlinks
\usepackage{url}            % simple URL typesetting
\usepackage{booktabs}       % professional-quality tables
\usepackage{amsfonts}       % blackboard math symbols
\usepackage{nicefrac}       % compact symbols for 1/2, etc.
\usepackage{microtype}      % microtypography
\usepackage{lipsum}
\usepackage{graphicx}
\graphicspath{ {./images/} }

\usepackage{cite}
\usepackage{amsmath,amssymb,amsfonts}
\usepackage{algorithmic}
\usepackage{graphicx}
\usepackage{textcomp}
\usepackage{url}
\usepackage{booktabs}

\title{A Data-Driven Biophysical Computational Model of Parkinson's Disease based on Marmoset Monkeys}

\author{
 Caetano M. Ranieri \\
 Institute of Mathematical and Computer Sciences \\
 University of Sao Paulo \\
 Sao Carlos, SP, Brazil \\
  \texttt{cmranieri@alumni.usp.br} \\
   \And
 Jhielson M. Pimentel \\
  Edinburgh Centre for Robotics\\
  Heriot-Watt University\\
  Edinburgh, Scotland, UK \\
  \texttt{jm210@hw.ac.uk} \\
  \And
 Marcelo R. Romano \\
  School of Electrical and Computer Engineering\\
  University of Campinas\\
  Campinas, SP, Brazil \\
  \texttt{marcelorromano@gmail.com} \\
  \And
 Leonardo A. Elias \\
  School of Electrical and Computer Engineering\\
  University of Campinas\\
  Campinas, SP, Brazil \\
  \texttt{leoelias@unicamp.br} \\
  \And
 Roseli A. F. Romero \\
 Institute of Mathematical and Computer Sciences \\
 University of Sao Paulo \\
 Sao Carlos, SP, Brazil \\
  \texttt{rafrance@icmc.usp.br} \\
  \And
 Michael A. Lones \\
  Edinburgh Centre for Robotics\\
  Heriot-Watt University\\
  Edinburgh, Scotland, UK \\
  \texttt{M.Lones@hw.ac.uk} \\
  \And
 Mariana F. P. Araujo \\
   Health Sciences Centre\\
  Federal University of Espirito Santo\\
  Vitoria, ES, Brazil \\
  \texttt{mfparaujo@gmail.com} \\
  \And
 Patricia A. Vargas \\
  Edinburgh Centre for Robotics\\
  Heriot-Watt University\\
  Edinburgh, Scotland, UK \\
  \texttt{p.a.vargas@hw.ac.uk} \\
  \And
  Renan C. Moioli \\
  Digital Metropolis Institute\\
  Federal University of Rio Grande do Norte\\
  Natal, RN, Brazil \\
  \texttt{renan.moioli@imd.ufrn.br}
}

\begin{document}
\maketitle
\begin{abstract}
In this work we propose a new biophysical computational model of brain regions relevant to Parkinson's Disease (PD) based on local field potential data collected from the brain of marmoset monkeys. Parkinson’s disease is a neurodegenerative disorder, linked to the death of dopaminergic neurons at the substantia nigra pars compacta, which affects the normal dynamics of the basal ganglia-thalamus-cortex (BG-T-C) neuronal circuit of the brain. Although there are multiple mechanisms underlying the disease, a complete description of those mechanisms and molecular pathogenesis are still missing, and there is still no cure. To address this gap, computational models that resemble neurobiological aspects found in animal models have been proposed. In our model, we performed a data-driven approach in which a set of biologically constrained parameters is optimised using differential evolution. Evolved models successfully resembled single-neuron mean firing rates and spectral signatures of local field potentials from healthy and parkinsonian marmoset brain data. {As far as we are concerned,} this is the first computational model of Parkinson's Disease based on simultaneous electrophysiological recordings {from seven brain regions of Marmoset monkeys.} Results show that the proposed model could facilitate the investigation of the mechanisms of PD and support the development of techniques that can indicate new therapies. It could also be applied to other computational neuroscience problems in which biological data could be used to fit multi-scale models of brain circuits.
\end{abstract}

% keywords can be removed

\section{Introduction}
\label{sec:introduction}

Parkinson’s disease (PD) affects more than 3\% of people over 65 years old, with figures set to double in the next 15 years~\cite{Poewe2017}. It is a neurodegenerative disease, whose symptoms include cognitive and motor deficits.  In late stages, it can also lead to depression and dementia~\cite{Shulman2011}. There is still no cure, and current therapies are only able to provide symptomatic relief. 

PD is characterised by a dopaminergic neuronal loss within the substantia nigra pars compacta (SNc), which leads to a dysfunction of the basal ganglia-thalamus-cortex (BG-T-C) circuit. The BG-T-C circuit is a neuronal network with parallel loops that are involved in motor control, cognition, and processing of rewards and emotions~\cite{Obeso2009TheObservations,Redgrave2010}.
{There are also links between the degeneration of dopamine neurons within those brain regions and changes in electrophysiological behaviour~\cite{galvan2008pathophysiology}}.

Brain regions linked to PD present complex interactions, with mutual excitatory and inhibitory feedback loops, which limit a comprehensive understanding of the physiopathology of the disease. Studies aiming at investigating the mechanisms underlying PD often use animal models. In classic animal models of PD, symptoms are elicited by delivering neurotoxins that damage the SNc dopaminergic neurons, such as 1-methyl-4-phenyl-1,2,3,6-tetrahydropyridine (MPTP) and 6-hydroxidopamine (6-OHDA), or chemicals that transiently inhibit dopamine production, such as alpha-methyl-p-tyrosine (AMPT)~\cite{Koprich2017AnimalDevelopment}. {Antipsychotics like haloperidol are also used, but have side effects that may promote dystonia and parkinsonism \cite{KHARKWAL2016}}.

To date, no animal model of PD fully reproduces human features of the disease. In addition, due to experimental limitations, animal data often include only a limited set of PD-related brain regions, with subjects engaged in different behavioural settings. In this context, computational models, with biologically informed constraints that can be selectively altered, are a promising, complementary approach to advance our knowledge about PD beyond that obtained from anatomical and physiological studies~\cite{Humphries2018,navarro2020dynamical}.
{Some PD-related anomalies observed in animal models, and efforts to reproduce those in computational models, are presented by Rubin \textit{et al.}~\cite{rubin2012basal}.}

All mammals have a similar set of BG structures that are similarly connected with thalamic and cortical structures. Nevertheless, recent studies suggest subtle differences between species~\cite{Lienard2014,Hardman2002,VanAlbada2009,Beul2017}, also in the neuropathophysiology of PD~\cite{Koprich2017AnimalDevelopment,Dawson2018}, with primates (including marmosets) being more similar to humans than rodents. For example, there are differences in the distribution of dopaminergic neurons in the substantia nigra of rats and primates, and the subthalamic nucleus and internal globus pallidus of rats have less neurons containing parvalbumin than primates \cite{Hardman2002}. Thus, a primate computational model of PD is of paramount importance.

In this work, we developed a new computational model of PD based on published data from the BG-T-C brain circuit of marmoset monkeys ~\cite{Cyranoski2009}. We built upon a neuronal computational model of rat models of PD, developed and made available by Kumaravelu \textit{et al.}~\cite{Kumaravelu2016ADisease}, and adjusted its parameters to match the electrophysiology data from the 6-OHDA+AMPT marmoset model of PD~\cite{Santana2014SpinalDisease,Santana2015}.

It is important to highlight that, in our work, we are using the LFP signal data to tune and validate our model, not spikes or other biosignals, thus the whole optimisation framework relates to LFP-based metrics. 
The main contributions of this paper are: (i) the first computational model of PD validated on simultaneous, multi-site electrophysiological recordings (e.g., LFP recordings) from a marmoset monkey model of the disease and (ii) an optimisation framework that can easily include novel biophysical parameters as soon as they become available.

This paper is organised as follows. In Section~\ref{sec:related-work}, models of the BG-T-C network and the anomalies caused by PD are discussed, with focus on computational modelling of the disease. In Section~\ref{sec:method}, the building blocks of the computational model are depicted, as well as the free parameters that were optimised, the algorithm to update those parameters, the experimental setup, and the evaluation protocol. In Section~\ref{sec:results-analysis}, the results are presented regarding the optimisation process, the parameters learnt by the machine learning algorithms, and the metrics observed on the simulations of the computational models provided, considering spectral densities from simulated LFP, dynamics of the firing rates from simulated neurons, and coherence analyses. In Section~\ref{sec:discussion}, a discussion is presented in order to contextualise our results and compare them with the expectations from the data from animal models, and knowledge from the literature. Section~\ref{sec:conclusion} concludes the paper.

\section{Related Work}
\label{sec:related-work}

\begin{figure*}[!ht]
    \centering
    \includegraphics[scale=0.6]{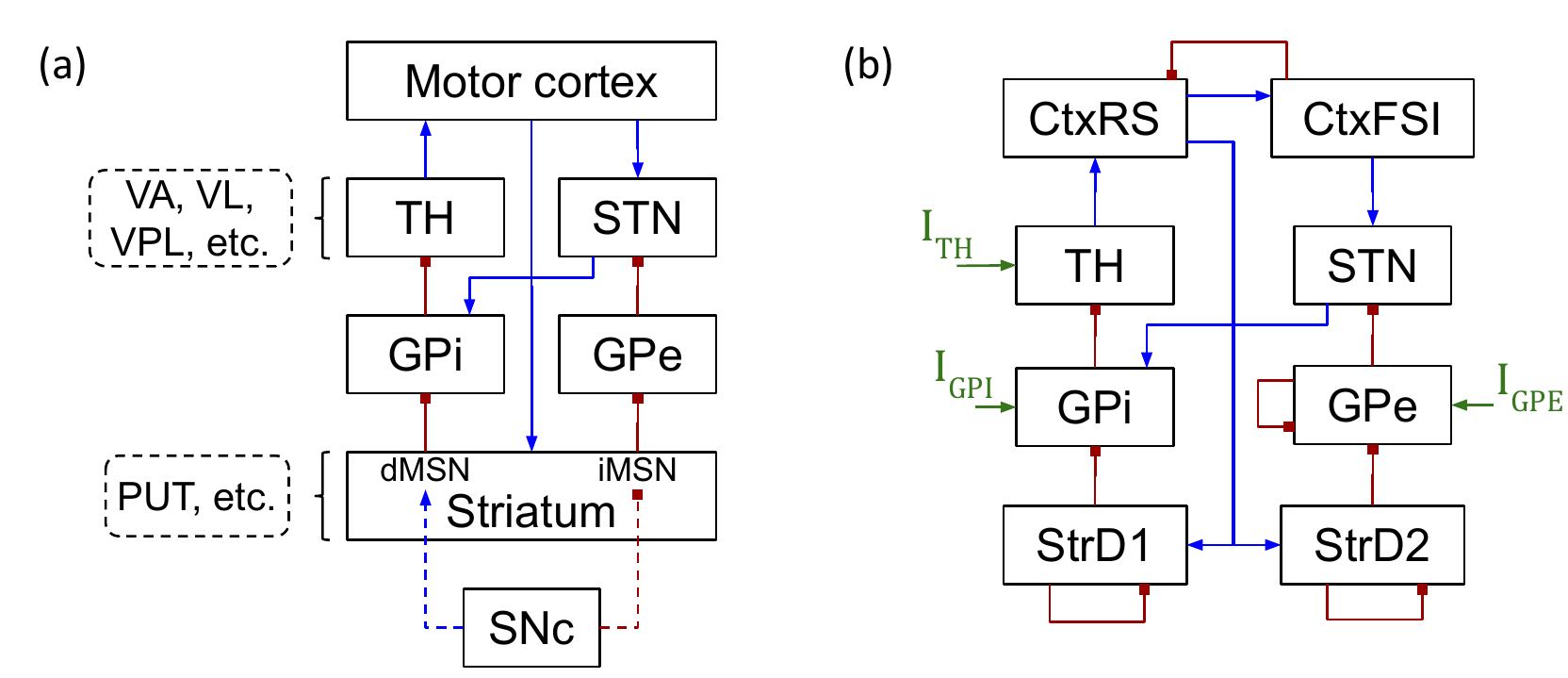}
    \caption{Models of the basal ganglia-thalamus-cortex (BG-T-C) circuit, central to the underlying mechanisms of Parkinson's Disease (PD), including excitatory (blue) and inhibitory (red) connections between the regions involved. (a) Classical model of BG-T-C circuit. The motor loop in the mammalian brain is formed by the \textit{motor cortex} (M1), the \textit{thalamus} (TH) --- composed of structures such as the \textit{ventral anterior nucleus} (VA), the \textit{ventral lateral nucleus} (VL), and the \textit{ventral posterolateral nucleus} (VPL) ---, and the \textit{basal ganglia} (BG), the latter composed of a subset of structures: the \textit{striatum}, which itself includes the \textit{putamen} (PUT) and the \textit{caudate nucleus}, the \textit{globus pallidus}, divided into \textit{pars interna} (GPi) and \textit{pars externa} (GPe), the \textit{subthalamic nucleus} (STN), and the \textit{substantia nigra}, divided into \textit{pars compacta} (SNc) and \textit{pars reticulata} (SNr). PD is caused by the loss of dopaminergic neurons in the SNc, which weakens the connections represented by dashed lines and leads to malfunctioning of both direct and indirect pathways. (b) BG-T-C network used in this work, based on \cite{Kumaravelu2016ADisease}. The cortex is represented by regular spiking (CtxRS) excitatory neurons and fast spiking (CtxFSI) inhibitory interneurons. The direct and indirect pathways in the striatum were modelled separately, representing the medium spiny neurons (MSNs) modulation by D1 and D2 dopamine receptors, respectively.}
    \label{fig:bg_circuit}
\end{figure*}

A commonly used model to explain how PD affects the neural connections within the BG-T-C circuit, also known as the motor loop, is the so-called classic model, illustrated in Figure~\ref{fig:bg_circuit}a. It consists of projections from primary motor (M1) and somatosensory cortical areas to BG input structures, specifically the putamen (PUT) and the subthalamic nucleus (STN). In PUT, the cortical projections establish excitatory glutamatergic synapses with medium spiny neurons (MSNs). 

The MSNs establish two distinct pathways to the BG output nuclei (globus pallidus pars interna – GPi and substantia nigra pars reticulata – SNr). The MSNs from the direct pathway (dMSN) directly project to the GPi/SNr, while the MSNs from the indirect pathway (iMSN) project to the globus pallidus pars externa (GPe), which in turn send projections to the GPi/SNr directly or indirectly via the STN (for reviews, see Obeso \textit{et al.}~\cite{Obeso2009TheObservations}, Lanciego \textit{et al.}~\cite{Lanciego2012}, and McGregor and Nelson~\cite{McGregor2019CircuitDisease}). 

The cortical projection to the STN establish a third pathway, often called the hyperdirect pathway \cite{Nambu2002}. Activation of the direct pathway facilitates movement by inhibiting the activity of GPi/SNr, thus reducing the inhibition of the ventral anterior nucleus (VA) and the ventral lateral nucleus (VL) and increasing the excitatory thalamic input to the motor cortex. Activation of the indirect and hyperdirect pathways, on the other hand, inhibit movement by increasing the inhibitory activity of the GPi/SNr over the VA/VL, hence decreasing the excitatory thalamic input to the motor cortex.

The activity of the motor loop is modulated by dopaminergic projections from SNc to PUT. The main effect of dopamine (DA) release in PUT is movement facilitation, since DA increases the excitability of the dMSNs and decreases the excitability of the iMSNs.
In PD, the depletion of striatal DA leads to an enhanced activation of the indirect pathway and a decreased activation of the direct pathway, resulting in the characteristic motor symptoms of this neural disorder \cite{Surmeier2014}. In addition to changes in firing rates, the functional imbalance within the motor loop in PD also disrupts the firing patterns within each nucleus and amongst the structures of the BG-T-C circuit, increasing neuronal synchronisation, neuronal bursting, and enhancing the oscillatory activity at the beta frequency band~\cite{Galvan2015}.

Bilateral 6-OHDA lesions in the marmoset medial forebrain bundle induce several PD motor symptoms, including impairments in fine motor skills, limb rigidity, bradykinesia, hypokinesia, and gait impairments.  Alpha-methyl-p-tyrosine (AMPT) administration to 6-OHDA lesioned marmosets can transiently increase the severity of these symptoms. However, like MPTP macaques, these animals do not exhibit resting tremor. Santana \textit{et al.}~\cite{Santana2015} provides an extensive characterisation of these symptoms, that were quantified through manual scoring (adapted version of the unified PD rating scale for marmosets), automated assessments of spontaneous motor activity in their home cages (using actimeters), and automated motion tracking while the animals explored two experimental apparatuses.

Computational models are established tools to facilitate understanding of neural disorders~\cite{Schroll2013,Sanger2018,Pena2020} and, in the context of PD, accommodate several levels of description and range from focusing on disease mechanisms to understanding anomalous neuronal synchronisation \cite{Humphries2018}. For instance, Pavlides \textit{et al.}~\cite{Pavlides2015} conducted a detailed study to help unveil the mechanisms underlying beta-band oscillations in PD and compared computational model predictions with experimental data. Muddapu \textit{et al.}~\cite{Muddapu2019} studied loss of dopaminergic cells in the SNc due to neural dynamics between SNc and STN, shedding light on the relevance of ongoing neural activity and neural loss. Gurney \textit{et al.}~\cite{Gurney2004} described mounting evidence relating the BG-T-C network and action selection mechanisms; actually, computational models showed a close relationship between action selection and BG-T-C oscillatory activity \cite{Humphries2006,Holgado2010,Merrison2017}.

{Moren \textit{et al.}~\cite{moren2019dynamics} proposed a model of the spiking neurons within the BG-T-C circuit, in order to observe the asynchronous firing rates around the 15 Hz beta-range oscillations, as well as on lower frequency bands.}
Terman \textit{et al.}~\cite{Terman2002} developed a conductance-based computational network model which shed light on the mechanisms underlying the neural dynamics of STN and GPe, a model which was further developed by Rubin \textit{et al.}~\cite{Rubin2004} to investigate the effects of deep brain stimulation (DBS) to eliminate anomalous synchronisation within the BG-T-C network in the PD condition. In fact, one of the key areas in which computational models serve as an invaluable tool for developing novel therapies is that related to predicting the effects of DBS~\cite{Humphries2012,Lu2020}.
Other computational models of the BG-T-C network and its relashionship with PD are presented by Farokhniaee and Lowery~\cite{farokhniaee2021cortical}, and Fleming \textit{et al.}~\cite{fleming2020simulation}.

Particularly important for the developments of this work is the research by Kumaravelu \textit{et al.}~\cite{Kumaravelu2016ADisease}, which, based on a collection of previously published studies, developed a computational model of the BG-T-C network tuned for the 6-OHDA \textit{rat model} of PD (Fig.~\ref{fig:bg_circuit}b). Compared to other computational models~\cite{Humphries2018}, it was the first to specifically consider 6-OHDA and a single species.

Most computational models related to BG-T-C dynamics rely on rodent data~\cite{Humphries2006,Lindahl2016,Koelman2019}, with only a handful focusing on primate data~\cite{Lienard2014,Topalidou2018}.
{Research by Shouno \textit{et al.}~\cite{shouno2017computational}, for instance, provided a spiking neuron model of the recurrent STN-GPe circuit for studying dysfunctions in oscillations within the 8-15 Hz (alpha) frequency band for PD primate models.}
In this work, we present a computational model to resemble neurophysiological activity of healthy and PD marmoset monkeys, based on the model by Kumaravelu \textit{et al.} \cite{Kumaravelu2016ADisease}, and in multisite, simultaneous LFP recordings from animal models.

\section{Methods}
\label{sec:method}

To provide a computational model of the BG-T-C circuit for PD-related features in primates, we began by re-writing the code by Kumaravelu \textit{et al.}~\cite{Kumaravelu2016ADisease}, originally implemented in Matlab.
We have ported the original code to the Python programming language, with the NetPyNE framework and the libraries from the NEURON simulator~\cite{Hines2009,Dura2019}.
Then, we performed a series of adaptations and employed a data-driven approach to calibrate a set of parameters, in order to derive a model that resembles local field potentials from marmoset data~\cite{Santana2014SpinalDisease,Santana2015}.
{More specifically, we employed an optimisation technique called differential evolution (DE), an algorithm based on evolutionary computation~\cite{ashlock2006evolutionary,corne2018evolutionary}.
This approach consists of optimising a predefined set of parameters (i.e., genotype) by gradually adapting them through successive steps (i.e., generations), providing variability and selection of the best solutions (i.e., individuals) through mechanisms analogous to biological evolution.}

{In the model by Kumaravelu \textit{et al.}~\cite{Kumaravelu2016ADisease}, neuronal connectivity and membrane initial conditions can be stochastic, and neuronal models include synaptic transmission delay. All currents (ionic, synaptic, leakage, and bias) are not subjected to noise. The model is general in the sense that it mimics the BG-T-C neural dynamics of rodents that are not engaged in any specific behavioural task. Thus, the dataset employed in our work was suitable to calibrate such a model, since it was collected from marmoset monkeys that were moving freely, without any event- or time-based stimulation.}

%{In the model by Kumaravelu \textit{et al.}~\cite{Kumaravelu2016ADisease}, no noise was introduced in the simulations. Neuronal connectivity and membrane initial conditions can be stochastic, and neuronal models include synaptic transmission delay. The dataset employed in our work was suitable to calibrate such a model, since it was collected from marmoset monkeys that were not engaged in any particular task, that is, they were moving freely, without any time-marked events such as sensory or artificial stimuli.}

After having calibrated our marmoset model, different analyses were performed in order to enhance and validate it.
The dataset used as ground truth for adjusting the parameters of the computational model is not publicly available due to legal restriction, but it is available from the corresponding author on reasonable request.
The next subsections will provide a detailed description of the methods employed.
The code to reproduce the results from this paper, including the machine learning framework and the analyses of the results, is publicly available at \url{https://github.com/cmranieri/MarmosetModel}.

\subsection{Computational Model}
\label{ssec:CM}

The computational model was based on Kumaravelu \textit{et al.}~\cite{Kumaravelu2016ADisease}. {Their model was build to reproduce the neurophysiological behaviour from rats based on data from healthy and 6-OHDA-lesioned individuals. As an initial step, we constructed an alternative implementation for their model within the NetPyNE framework, and we validated this implementation by comparing its outputs with those reported in \cite{Kumaravelu2016ADisease}.}

Briefly, eight brain structures were modelled and connected based on a simplified version of the classic model (Figure~\ref{fig:bg_circuit}b). In particular, the direct and indirect pathway in the striatum were modelled separately representing the MSN modulation by D1 and D2 dopamine receptors, respectively~\cite{McGregor2019CircuitDisease}. The cortex is represented by regular spiking (RS) excitatory neurons and fast spiking (FSI) inhibitory interneurons. Neurons from all but cortical regions were modelled using a biophysically based Hodgkin–Huxley~\cite{hodgkin1952currents} single-compartment model, whereas cortical neurons were constructed based on the computationally efficient Izhikevich's model~\cite{Izhikevich2003}. The reasoning for different neuronal models lies on the fact that PD effects are captured by altering specific conductances in selected structures (see below), thus a con\-duc\-tance-based model is more suitable at these locations. Finally, a bias current was added in the TH, GPe, and GPi, accounting for the inputs not explicitly modelled. {Remarkably, even though no oscillatory inputs are present in the model, synaptic delays and network interactions by means of recurrent connections promote sustained firing rate oscillations.} For a detailed description of connectivity schemes and other implementation details, the reader is referred to Kumaravelu \textit{et al.}~\cite{Kumaravelu2016ADisease}.

The computational model described above can shift from the simulations of the healthy to the PD condition by altering three conductances~\cite{Kumaravelu2016ADisease}: decreasing the maximal M-type potassium conductance in direct and indirect MSN neurons (MSN firing disfunction) from 2.6 to 1.5 $mS/cm^{2}$; decreasing the maximal corticostriatal synaptic conductance (reduced sensitivity of direct MSN to cortical inputs) from 0.07 to 0.026 $mS/cm^{2}$; and increasing the maximal GPe axonal collaterals synaptic conductance from 0.125 to 0.5 $mS/cm^{2}$ (increase of GPe neuronal firing). This is implemented in the model with a control flag.

One major addition to the model developed here is the simulation of local field potentials (LFP). {These measurements are related to the extracellular activity produced by action potentials of the neurons within a brain region~\cite{gold2006origin}. A discussion on the dynamics of LFP signals within the basal ganglia and its consequences to humans, especially regarding conditions such as PD, was presented by Brown and Williams~\cite{brown2005basal}.}

{In our work, first, each simulated brain region is assigned to a spatial 3D coordinate that matches that used in the stereotaxic surgery where electrodes were placed in the real marmoset monkeys~\cite{Paxinos2012,Santana2014SpinalDisease}. Then, a simulated electrode is placed at the centre of each region. In our model, each neuron is represented as a single cylindrical compartment with a membrane area of 100 $\mu m^{2}$. For each electrode, NetPyNE estimates the simulated LFP by summing the extracellular potential contributed by each neuronal segment (based on the transmembrane current generated from the single cylindrical source neuron), calculated using the “line source approximation” method and assuming an Ohmic extracellular medium with conductivity $\sigma=0.30$ $mS/mm$~\cite{Parasuram2016}. Thus, the electrical activity of neurons from each brain region contributes to the waveforms recorded at each electrode (subject to extracellular medium attenuation).}

\subsection{Dataset and preprocessing procedures}
\label{ssec:data-preprocessing}

{The dataset we used in the present work is based on a previous study by Santana \textit{et al.} \cite{Santana2014SpinalDisease}. Our dataset includes data from three adult males and one adult female common marmosets (i.e., \textit{Callithrix jacchus}). Data from two males were part of the aforementioned study; data from one male and one female are novel and followed exactly the same experimental procedures. A short summary is presented in the next subsection, followed by the preprocessing steps.}

\subsubsection{Dataset}
\label{sssec:dataset}

The animals, weighing 300–550 g, were housed in a vivarium with natural light cycle (12/12 hr) and outdoor temperature. All animal procedures followed approved ethics committee protocols (CEUA-AASDAP 08/2011, 11/2011 and 03/2015) strictly in accordance with the National Institutes of Health (NIH) Guide for the Care and Use of Laboratory Animals. PD symptoms were elicited in all three male animals with injections of 6-OHDA toxin in the medial forebrain bundle under deep anaesthesia \cite{Santana2015,Santana2014SpinalDisease}. Prior to neural recordings, animals that received 6-OHDA were subjected to acute pharmacological inhibition of dopamine synthesis (subcutaneous injections of AMPT $2\times3 240$ mg/kg) to further exacerbate PD motor symptoms, mimicking a more severe stage of the disease. Although 6-OHDA lesions impact on both behavioural and electrophysiological features in all animals \cite{Santana2015,Santana2014SpinalDisease}, there are individual differences at earlier stages of dopaminergic depletion that could hinder our model development considering the relatively low number of subjects.

Both healthy and PD animals were implanted each with two custom-made microelectrode arrays composed of 32 microwires (one array in each hemisphere). The wires were 50 $\mu m$ in diameter and were organised in bundles aimed to reach distinct areas of the BG-T-C system. Before the surgery, the animals were sedated with ketamine (10-20 mg/kg i.m.) and atropine (0.05 mg/kg i.m.), followed by deep anesthesia with isoflurane 1-5\% in oxygen at 1-1.5 L/min. The arrays were then implanted  using a stereotaxic manipulator to position electrodes at the targeted BG-T-C coordinates, which were determined using Stephan \textit{et al.}~\cite{Stephan1980} and Paxinos \textit{et al.} \cite{Paxinos2012} stereotaxic atlas. The microelectrode array and the implant procedures were thoroughly described in Budoff \textit{et al.}~\cite{Budoff2019}.

Once the animals recovered from the surgery, recording sessions were performed in fully awaken animals behaving freely.  LFPs were sampled at 1,000 Hz and recorded using a 64 multi-channel recording system (Plexon). The position of the recording microelectrodes were verified postmortem through either tyrosine hydroxylase (TH) staining or Nissl staining. Similarly, the extent of dopaminergic lesions were verified through the quantification of striatal fiber density and dopaminergic midbrain cells in TH-stained sections. Further experimental details are described in Santana \textit{et al.}~\cite{Santana2014SpinalDisease}.

\subsubsection{Preprocessing}
\label{ssec:preprocessing}

For our study, in total, 14 and 16 recording sessions were taken for the healthy and PD conditions, respectively, considering the brain hemispheres independent from each other. For the PD condition, we recorded from M1, PUT, GPe, GPi, ventral lateral (VL) and ventral posterolateral (VPL) thalamic nuclei, and STN, whereas for the healthy animal regions M1, PUT, GPe, and GPi were recorded. The raw data was organised so that, for each recording session, a data structure with $N_\text{elec} \times N_T$ was provided, where $N_\text{elec}$ is the number of electrodes recorded and $N_T$ is the number of samples of the recording session (variable but typically lasting for several minutes).

%To employ data-driven strategies based on machine learning, a more uniform data structure was needed, and this requirement was taken into account when developing the preprocessing pipeline.}

\begin{figure*}
    \centering
    \includegraphics[scale=0.4]{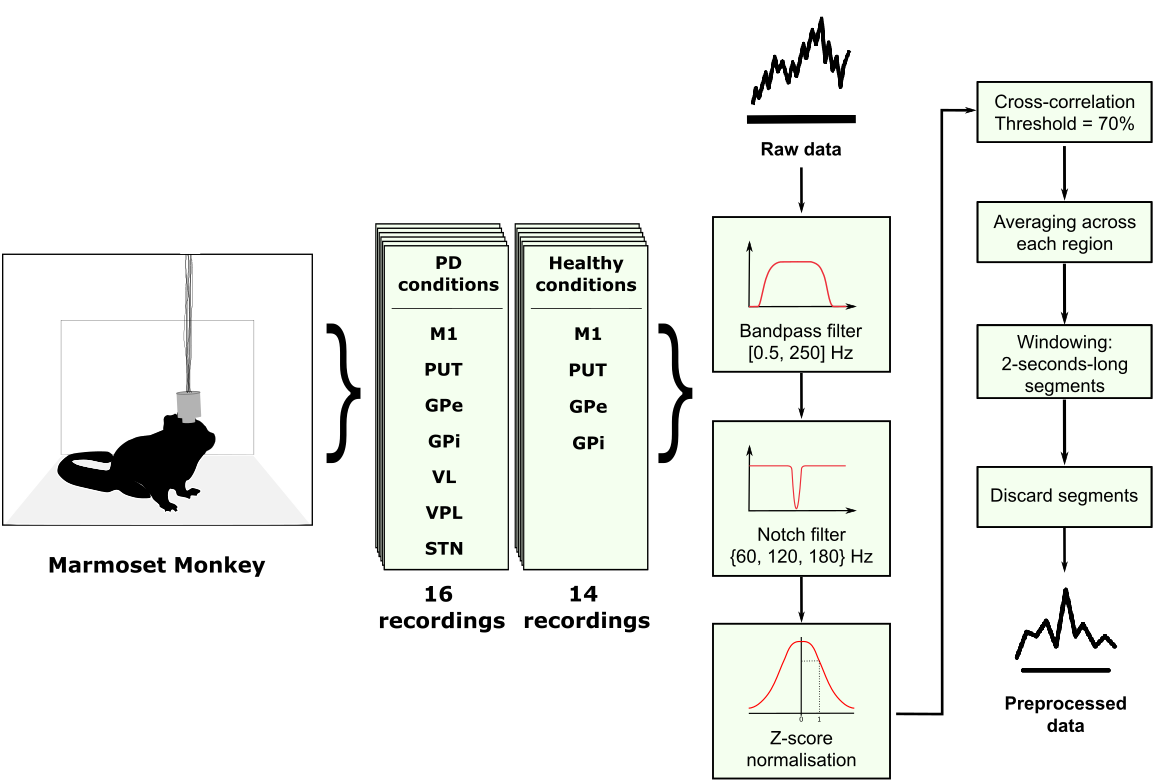}
    \caption{Data acquisition and preprocessing steps implemented in out method. Depending on the monkey condition (healthy or PD), different regions of the brain were recorded. The input data was composed of a whole recording session, with variable lengths and numbers of channels (i.e., electrodes) per region. After preprocessing, the data was transformed into 2-seconds-long segments with seven channels, each related to one of the regions analysed.}
    \label{fig:preprocessing}
\end{figure*}

Figure~\ref{fig:preprocessing} illustrates the preprocessing steps adopted after data acquisition. For each channel, the pipeline began with a zero-lag low-pass filter (cutoff frequency of 250 Hz) and a high-pass filter (cutoff frequency of 0.50 Hz), to eliminate frequencies that are outside the LFP scope and may relate to electrical or mechanical interference. Then, we minimised power grid interference (hum) with a notch filter centred at 60 Hz and its harmonics (120 Hz and 180 Hz). Each resulting signal was then scaled according to a z-score normalisation, to account for the possible differences in signal amplitude due to different electrode impedance.

%we grouped the readings per region (e.g., readings from the M1 were considered together), with the aim to provide a single row of data for each of the regions considered. Then,

In the next step, we computed the cross-correlation matrix $Q$ according to Equation~\ref{eq:crosscoef}, where $C_{ij}$ is the covariance matrix of the filtered and z-scored signals from electrodes $i$ and $j$, which are located exclusively within a brain region. Channels within each region with mean correlation coefficient below the threshold of 0.70 were discarded.
This procedure was employed because electrodes in each recorded region are placed very close to each other (see electrode and surgical procedures above), thus we expect LFP signals to be highly correlated (if they are not, it may relate to a noisy electrode signal) \cite{Buzsaki2012TheSpikes.}.

\begin{equation}
    \label{eq:crosscoef}
    Q_{ij} = \frac{ C_{ij} } { \sqrt{C_{ii} \cdot C_{jj}} }
\end{equation}

All remaining LFP channels within a brain region were averaged, which provided one data matrix for each recording session with dimensions $N_R \times N_T$, where $N_R$ is the number of brain regions recorded.
{These average LFP values were computed based solely on channels within each region.}
Next, we segmented each time-series in 2-second segments, which was the same length as the computational model simulations (see Section~\ref{ssec:evolutionary-strategy} for details).
Considering the data sampling rate (1,000 Hz) and frequencies of interest (up to 50 Hz), 2-second segments provide enough data for our analyses. Prohibitively noisy segments were discarded using two criteria: first, segments with abnormal amplitudes, detected using an upper threshold of 0.20 for the absolute value of the mean of the signal over time; second, segments with limited (abnormal) oscillatory patterns, detected using a lower threshold of 0.10 for the amplitude standard deviation and a minimum threshold of 10 amplitude peaks. For each recording session, our preprocessed dataset had a final shape of $N_R \times 2000 \times N_\text{seg}$, where $N_\text{seg}$ is the resulting number of segments.

In the dataset adopted for this work, whether animals were still or moving could have a profound effect on brain oscillatory activity and synchronisation metrics, because all animals were behaving freely and were not engaged in any particular behavioural task during the recording sessions.
In fact, especially in motor and pre-motor regions, modulations in neural oscillatory dynamics linked to motor activity are well characterised (see Armstrong \textit{et al.}~\cite{Armstrong2018} for a review), and recent studies show that even breathing can modulate neural oscillations \cite{Tort2020}. However, we understand that action initiation, movement, or breathing have low influence on averaged LFP amplitude values computed, given that the 2-second window segments were randomly selected without time alignment to any specific movement or action.

\subsection{Evolutionary Algorithm}
\label{ssec:evolutionary-strategy}
%https://link.springer.com/article/10.1007/s00521-020-04832-8

{Evolutionary algorithms are optimisation techniques in which a set of parameters, called \textit{genotypes}, are gradually recombined and changed according to mechanisms analogous to those of biological evolution, in order to maximise a fitness function dependent of those parameters \cite{ashlock2006evolutionary}.}
Differential evolution (DE)~\cite{RAINERSTORN1997} was employed to fit the computational model parameters so that it matches the LFP beta-band power spectrum observed in the marmoset data.

{The overall structure of the model was preserved from Kumaravelu \textit{et al.}~\cite{Kumaravelu2016ADisease}, while a set of conductances, background currents and synaptic modulations, as well as the numbers of neurons in each region of the BG-T-C circuit, were calibrated through the evolutionary algorithm.
The connectivity, the delays, the synaptic mechanisms, the remaining conductances, and all other parameters were kept as in the original model (see Section 3 from the Supplementary Material).}

{More specifically}, fourteen parameters compose the {set of parameters to be optimised (i.e., the genotype)}. Parameter $I_{TH}$ ($\mu A/cm^2$) relates to cerebellar input currents to the thalamus, which are linked to sensorimotor inputs~\cite{Manto2013}. Parameters $I_{GPe}$ ($\mu A/cm^2$) and $I_{GPi}$ ($\mu A/cm^2$) relate to currents at GPe and GPi, respectively, from all sources that were not explicitly modelled. The next two parameters, $g_{STN\_KCA}$ (nS/cm\textsuperscript{2}) and $g_{GP\_AHP}$ (nS/cm\textsuperscript{2}), refer to the maximum slow potassium conductance yielding afterhyperpolarization (AHP) at the STN and the calcium-activated potassium conductance at GPe and GPi, respectively. The sixth parameter, $g_{syn\_CTX\_STR}$ (nS/cm\textsuperscript{2}), modifies the synaptic conductance from cortex (CTX) to striatum (STR). Finally, parameters seven to 14 map to the number of neurons in each modelled region. All of the aforementioned parameters were chosen because they have a direct influence on the firing rates of neurons within each region, which in turn affect the LFP \cite{Parasuram2016}. Also, comparing marmoset with rodent literature, there is very limited quantitative work on the anatomical and neurophysiological parameters of the BG-T-C neuronal network.

\begin{table}[t]
\centering
\caption{Free parameters of the computational model, optimised by DE to fit the marmoset data.}
\label{table:params}
\vspace{6pt}
\begin{tabular}{@{}llll@{}}
\toprule
ID & Parameter & Range & Description \\ \midrule
1 & $I_\text{TH}$ & $[0.6, 1.8]$ & Background currents at TH ($\mu A/cm^2$) \\
2 & $I_\text{GPe}$ & $[1.5, 4.5]$ & Background currents at GPe ($\mu A/cm^2$) \\
3 & $I_\text{GPi}$ & $[1.5, 4.5]$ & Background currents at GPi ($\mu A/cm^2$) \\
4 & $\bar{g_\text{STN\_KCa}}$ & $[2.5,7.5]$ & $Ca^{2+}$--dependent AHP $K^{+}$ conductance \\ & & & at STN ($mS/cm^2$) \\
5 & $g_\text{GP\_AHP}$ & $[5.0,15.0]$ & $Ca^{2+}$--dependent AHP $K^{+}$ conductance \\ & & & at GPe and GPi ($mS/cm^2$) \\
6 & $g_\text{syn\_ctx\_str}$ & $[0.8,1.2]$ & Synaptic modulation from cortex to \\ & & & striatum ($mS/cm^2$) \\
7 & $n_\text{GPe}$ & $[10,30]$ & Number of GPe neurons \\
8 & $n_\text{GPi}$ & $[10,30]$ & Number of GPi neurons \\
9 & $n_\text{TH}$ & $[10,30]$ & Number of TH neurons \\
10 & $n_\text{StrD1}$ & $[10,30]$ & Number of StrD1 neurons \\
11 & $n_\text{StrD2}$ & $[10,30]$ & Number of StrD2 neurons \\
12 & $n_\text{CTX\_RS}$ & $[10,30]$ & Number of CTX\_RS neurons \\
13 & $n_\text{CTX\_FSI}$ & $[10,30]$ & Number of CTX\_FSI neurons \\
14 & $n_\text{STN}$ & $[10,30]$ & Number of STN neurons \\ \bottomrule
\end{tabular}
\end{table}

In the DE, each individual from the population was a model $M(G)$ that consisted of an adaptation of the model of Kumaravelu \textit{et al.}~\cite{Kumaravelu2016ADisease} in the PD condition, in which the parameters of Table~\ref{table:params} were set to the values defined by genotype $G$. Each model $M(G)$ was simulated for $t_{sim}=2000$ milliseconds, and the spike trains from each neuron and LFPs from each virtual electrode were recorded.
The LFP recordings were applied to calculate the fitness function $f(M)$ as follows.

Given a categorical set $R$ containing $N_R$ brain regions, the mean power spectral density (PSD) of the LFP from the electrode placed in region $r \in R$ is denoted by $S_r$ and defined in  Equation~\ref{eq:avg-psd}, where $[ \omega_a, \omega_b ]$ is the frequency interval of interest and $\hat{P}_r(\omega)$ is the periodogram computed with Welch's method~\cite{Rao2018SpectralSignals}.

\begin{equation}
    \label{eq:avg-psd}
    S_r( \omega_a, \omega_b ) = \int_{\omega_a}^{\omega_b} \hat{P}_r(\omega) d\omega
\end{equation}

According to the literature on the electrophysiology of PD~\cite{Poewe2017,Tinkhauser2017BetaMedication}, a noticeable abnormality is observed typically at the centre of the beta frequency band of LFP recordings from the basal ganglia of PD individuals.
This frequency band corresponds approximately to the interval [13,30] Hz, although this range varies within human patients and animal model species.
For the formulation of the fitness function, let a coefficient $y_r$ be the summation of the beta-band mean PSD plus the mean PSD of adjacent bands, composing the interval [8,50] Hz, normalised by the mean PSD of all frequencies up to $50$ Hz, as stated in  Equation~\ref{eq:y_r}.
This broader interval was defined to account for possible wider spectrum modulations in adjacent bands.

\begin{equation}
    \label{eq:y_r}
     y_r = \frac{S_r(8, 50)}{S_r(0.5, 50)}
\end{equation}

The fitness function $f(M)$ is defined in  Equation~\ref{eq:fitness}, where $y_{r(target)}$ is the average value of  Equation~\ref{eq:y_r} calculated from the preprocessed data of all marmosets of PD condition, and $y_{r(M)}$ is calculated considering the simulated LFP of a computational model $M$. Notice that the healthy marmoset condition lacks readings from TH and STN regions (i.e., no electrodes were implanted in these regions). In addition, the dataset includes three PD model animals. For this reason, DE optimised parameters for mimicking the PD condition. Fitness values vary from 0, if simulated and marmoset data LFP in all brain regions substantially differ, to $N_{R}$, if they match.

\begin{equation}
    \label{eq:fitness}
    f(M) = N_R - \sum_{r \in R} \min \left\{ 1, \left| \frac{y_r(M) - y_r(target)}{y_r(target)} \right| \right\}
\end{equation}

Eight brain regions are simulated, thus $N_{R}=8$. PSD target values for the simulated regions StrD1 and StrD2 are drawn from marmoset LFP PSD values for PUT. Simulated TH is tuned based on the average PSD from marmoset VL and VPL, and simulated CtxRS and CTxFSI are tuned based on marmoset M1. Simulated GPe, GPi, and STN LFP PSDs are matched to the respective marmoset LFP PSDs.

%The GA would be expected to find a genotype $G = G^\text{*}$ that maximises $\mathcal{f}[M(G)]$, which outputs values in the range $[0,N_R]$. This would be defined as the best individual of the final population of the algorithm. 

The DE initial population was set to 200 individuals, whose initial parameters were drawn from a random uniform distribution in the interval $[0, 1]$. Parameters were normalised to the ranges listed in Table~\ref{table:params} (i.e., the actual values set in the computational model) only at simulation time. In each DE generation, a set of 20 individuals were selected through tournaments of size two. Pairs of those selected individuals were randomly chosen, in order to generate two offspring by applying uniform crossover. This led to a child population of size 20. The mutation rate was set to 10\% and followed a normal distribution $\mathcal{N}(\mu=0.0,\sigma=0.3)$. The DE implements generational replacement with elitism, with only one elite individual of the parent population being kept, resulting in a population size of 21 individuals. Each model $M(G_k)$, where $k \in \{ 1, \dots, N_M \}$, was evolved for $N_{gen}=60$ generations. We have performed 150 evolutionary runs, so that the highest fitness individual of each run was selected to compose the set $\mathcal{G}=\{ G_1, \dots, G_{N_M} \}$ of evolved genotypes.

\subsection{Clustering}
\label{ssec:clustering-analysis}

Upon completion of parameter optimisation by DE, we investigated whether high fitness individuals had different genotypes. The rationale is that different parameter sets, even if biologically plausible, could lead to incompatible healthy and PD network dynamics \cite{Bahuguna2017}. 
Considering that the fitness function was computed based on LFP values of the PD condition only, and that the healthy condition was obtained by changing the same parameters listed by Kumaravelu \textit{et al.}~\cite{Kumaravelu2016ADisease}, there was no guarantee that the genotypes evolved would lead necessarily to models that resemble the healthy and PD conditions of the animal models.
For this reason, we performed a clustering analysis~\cite{Verma2012AMining} to the set $\mathcal{G}$ of evolved genotypes, which we could then evaluate separately based on their spectral densities. This validation step is based on the fact that PD individuals present a peak at the beta band (13-30 Hz) when compared to healthy individuals~\cite{Tinkhauser2017BetaMedication}.

Let $\mathcal{C}=\{C_1,\dots,C_{n_c}\}$ be a set of clusters, with $C_p=\{G_1,\dots,G_{n_p} \}$, where $n_c$ is the number of clusters, $p \in \{1,\dots,n_p\}$, and $n_p$ is the total number of genotypes within cluster $p$.
Considering $s_p(G_k)$ to be the sample silhouette \cite{rousseeuw1987silhouettes} of genotype $G_k$ with respect to $C_p \in \mathcal{C}$, consider $s_p(G_k) \geq s_p(G_{k+1})$ for all $k \in [1,P]$, it is, each cluster is ordered from highest to lowest silhouette.
In exploratory experiments (not shown), we investigated different clustering paradigms, namely K-means, density-based spatial clustering of applications with noise (DBSCAN), and agglomerative clustering. Based on these experiments, we opted for the K-means algorithm with two centroids (i.e., $p=2$), because this configuration led to the highest mean silhouette score. Hence, the K-means algorithm was fed with all the individuals with the highest fitness per evolutionary run (i.e., set $\mathcal{G}$), and the Euclidean distances for the algorithm were computed on the 14 normalised parameters of the genotype.

\subsection{Computational model spike and LFP analysis}
\label{ssec:spikeLFP-analysis}

The different clusters of genotypes were compared with respect to their parameter values, spike firing rates and LFP power spectra. For each cluster, the 50 highest fitness genotypes were chosen for the following analyses.
Spectral analysis was performed by simulating $C_p[1,\dots,50]$, for $t_{sim}=2000$ milliseconds, in both healthy and PD conditions. Thus, for each condition, $50$ simulated LFP recordings were analysed per cluster for each condition.
Since we simulated the same individuals (i.e., sets of parameters), with the same seeds for generation of random numbers, in each of the conditions (healthy and PD), the samples across these conditions were considered to be dependent. The PSDs were computed and evaluated with respect to the mean of the density spectrum per cluster, and the average power at the beta band.

For PSD analyses on the LFP of either the animal and computational models, to highlight the presence of a peak in the beta band in the PD condition, a ratio $R$ was defined as in Equation \ref{eq:r}, where $\hat{P}_{r}^{PD}(\omega)$ and $\hat{P}_{r}^{H}(\omega)$ are the mean spectral power across the PD and healthy models, respectively, for frequency $\omega$. A lower threshold value $\epsilon$ was defined because, for denominators too close to zero, the ratio may lead to high values that actually have little meaning for interpretation. For the analyses with the animal models, $\epsilon$ was defined as the median power across the mean spectrum of the healthy condition. For the computational models, it was set to the $80$th percentile of the healthy spectrum.

\begin{equation}
\label{eq:r}
R(\omega) =
\begin{cases}
    \frac{ \hat{P}_{r}^{PD}(\omega)}{ \hat{P}_{r}^{H}(\omega) } \quad , \quad \hat{P}_{r}^{H}(\omega) > \epsilon \\
    0 \quad , \quad \text{otherwise}
\end{cases}
\end{equation}

Regarding spike dynamics, the models within each cluster were simulated for $t_{sim}=2000$ milliseconds with time step size $dt = 0.10$ milliseconds, always with the same seed for random number generation, and the firing frequency of all neurons was calculated in $50$ time bins, each corresponding to $20$ milliseconds. 

\subsection{Computational model validation}
\label{ssec:model-validation}

Considering that different currents, conductances, and numbers of neurons may influence the firing rate in each simulated brain region, which in turn modulates the LFP power spectra, one may conclude that even if there are different clusters, their neural dynamics are comparable because both clusters are formed by high fitness individuals. However, even if our computational model was optimised to replicate the LFP power spectra from marmoset animal models of PD, it should also mimic the power spectra from healthy marmosets (by changing selected conductances, see Section~\ref{ssec:CM}). In other words, if the computational model accurately captures the physiological phenomena responsible for the different beta-band centred LFP power spectra from PD marmoset monkeys, it should also replicate the healthy spectra (a scenario in which it was not evolved).

Therefore, we first confirmed that our marmoset animal model of PD presented frequency spectra in accordance with previous works, following Section~\ref{ssec:data-preprocessing}. Then, we investigated whether the computational model would also capture this phenomena. For that, for each genotype cluster found (Section~\ref{ssec:clustering-analysis}), we compared the LFP power spectra from the evolved PD computational model with that from the healthy model. This was performed by modifying a predefined set of conductances in the simulation (Section~\ref{ssec:CM}). To highlight the differences, we first analysed the ratios between the mean PSD of the PD and healthy simulated individuals from each cluster. 

During evolution, fitness is given by LFP PSD in the vicinity of the beta band calculated in the whole $t_{sim}=2000 ms$ sequence, hence it is possible that the same spectra relate to different LFP rhythms over shorter time scales. Thus, different neuronal spiking dynamics may lead to similar LFP dynamics over time. Moreover, spikes from single neurons are noisy and vary considerably over time and over repeated simulations. With large recordings, joint neuronal averages over time may hinder comprehension of neural population dynamics. Finally, one of the advantages of computational models such as the one used here is the direct access of each neuron state at any given time, but it is not trivial to interpret the dynamics of large populations of neurons over time. To clarify these issues, we studied low-dimensional neuronal trajectories for both healthy and PD computational model conditions~\cite{Cunningham2014}.

To compute the neuronal trajectories, we first calculated the firing frequencies for all neurons from each simulated model in a particular cluster and condition (i.e., healthy and PD), based on the mean firing rates (MFR) taken from bins of size $50$ ms. Since the number of neurons within each region varies from 10 to 30 (see Table~\ref{table:params}), and there are eight regions considered for the computational model, this procedure generates time series with high dimensionality, ranging from 80 to 240, which would be difficult to visualise and analyse. To reduce the dimensionality, we employed principal component analysis (PCA) \cite{wold1987principal}; that is, we analysed neural trajectories by projecting high-dimensional neural population activity in a 3D space using PCA of the spike MFR time series.

However, what if, instead of clearly occupying different regions in the state space, neuronal responses from the same conditions result in similar paths in the reduced dimensional space? To address this hypothesis and to compare PCA trajectories, we used Dynamic Time Warping (DTW) with Euclidean distance \cite{Muller2007}.
DTW finds the optimum non-linear alignment between two time series, hence it can estimate whether neuronal trajectories share a similar path, regardless of initial conditions.
In the analysis performed, we employed the \textit{fastdtw} Python package, which implements the method proposed by Salvador and Chan~\cite{salvador2007toward}. Each pair of three-dimensional time series, computed from the MFR signals and dimensionally reduced with PCA, was fed to the algorithm, which provided, as output, a scalar proportional to the dissimilarity between the two time series being compared.

More specifically, we compared the similarity of all possible pairs of neural trajectories considering all individuals within the clusters (healthy and PD dynamics). We compared all pairs of trajectories generated by individuals within the same condition (healthy or PD), which gave a measurement of how different the healthy or PD individuals are compared to each other (i.e., within-group comparison), and we compared pairs of trajectories between healthy and PD conditions (i.e., between-groups comparison).

Finally, one of the hallmarks of PD is the anomalous widespread synchronisation in the BG-T-C network. To validate our model in that respect, we calculated the magnitude-squared coherence between nuclei and intranucleus. Based on a similar analysis performed in healthy and PD marmosets reported in Santana \textit{et al.}~\cite{Santana2014SpinalDisease}, we expect a widespread increase in this metric. The magnitude-squared coherence was calculated from the spike trains of neurons of each nucleus using Welch's method with Hanning windowing without overlap and with spectral resolution of 1 Hz. The average was taken as recommended by Bendat and Piersol~\cite{Bendat2010}: the squared value of the average of the cross spectra divided by the product of the mean values of the auto spectra of each nucleus.

The value of the magnitude-squared coherence between brain regions $r_A$ and $r_B$, defined as $C(r_A,r_B)$, was computed as in Equation~\ref{eq:coh}, where $N_A$ is the number of neurons in region $r_A$, and $N_B$ is the number of neurons in region $r_B$, and $S(r_x^{m},r_y^{n})$ is the cross spectrum between the spike trains from the $m$-th neuron from region $r_x$ and the $n$-th neuron from region $r_y$.

\begin{equation}
    \label{eq:coh}
    C(r_A,r_B)=\frac{ \left[ \frac{1}{N_A \cdot N_B} \sum\limits_{i=1}^{N_A} \sum\limits_{j=1}^{N_B} S(r_A^{i},r_B^{j}) \right] ^2}
    {\left[ \frac{1}{N_A}\sum\limits_{i=1}^{N_A} S(r_A^{i},r_A^{i}) \right] \cdot
    \left[ \frac{1}{N_B}\sum\limits_{i=1}^{N_B} S(r_B^{i},r_B^{i}) \right]}
\end{equation}

Then, we considered the peak of the coherence in the 7-30 Hz band to highlight PD-related effects~\cite{Santana2014SpinalDisease}.
The significance level for coherence was defined as $1-(1-\alpha)^{1/(L-1)}$~\cite{Rosenberg1989}, with $\alpha = 0.95$ and $L = 100$, because the windowing was done with 100 segments and we adopted as 95\% the significance level. As the computational models have eight nuclei, an $8\times8$ matrix was constructed, representing the coherence between each pairs of nuclei.
The median of this matrix was considered as the global coupling metric between nuclei in each simulation, because it is less sensitive to outliers than the mean.

\section{Results}
\label{sec:results-analysis}

%p-value annotation legend:
%ns: 5.00e-02 < p <= 1.00e+00
%*: 1.00e-02 < p <= 5.00e-02
%**: 1.00e-03 < p <= 1.00e-02
%***: 1.00e-04 < p <= 1.00e-03
%****: p <= 1.00e-04

\begin{figure*}[!ht]
        \centering
        \includegraphics[scale=0.75]{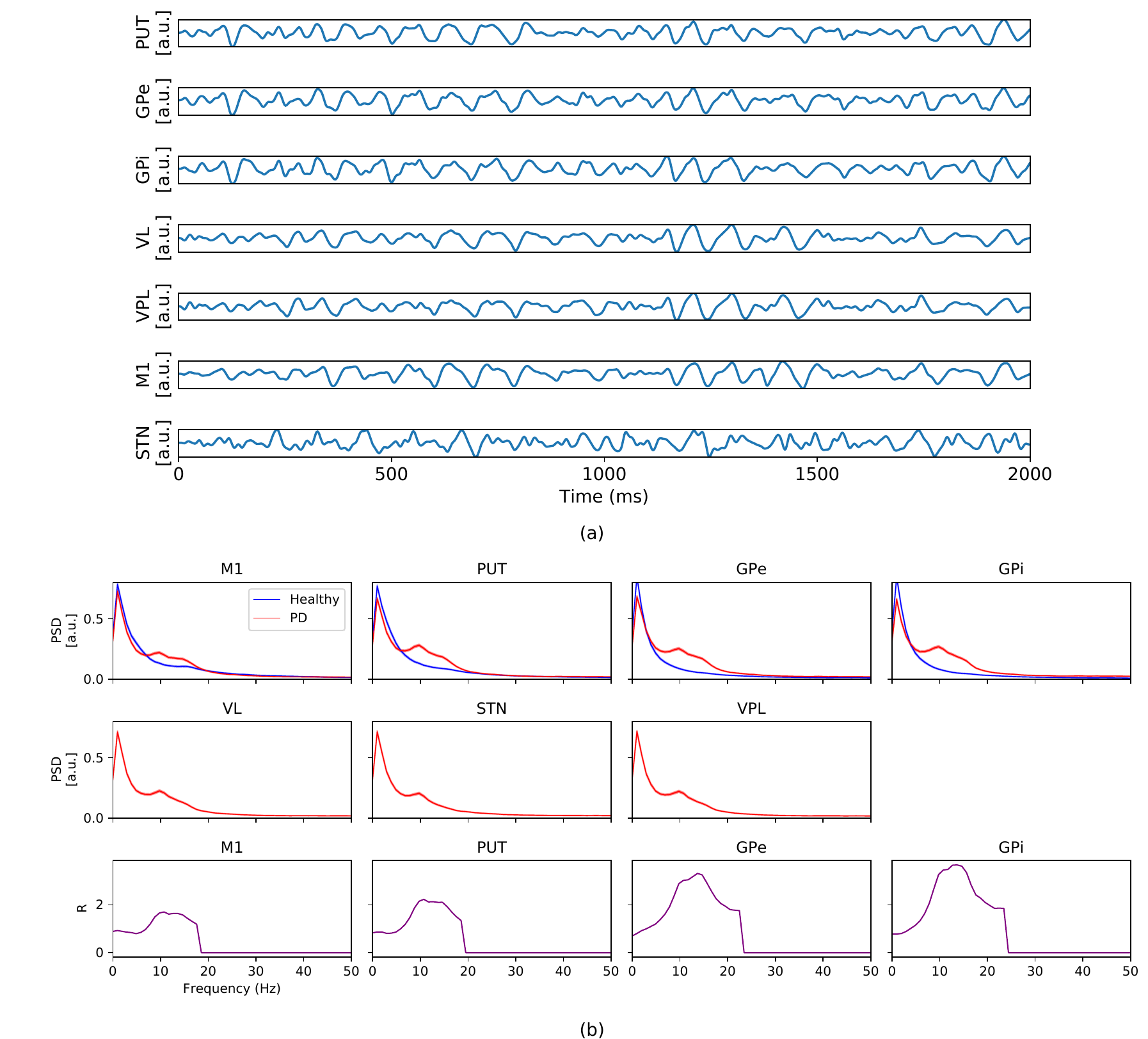}
        \caption{Animal data from marmoset monkeys, collected through electrodes implanted to each region of the BG-T-C circuit in a previous study \cite{Santana2014SpinalDisease}, and made available for our research. (a) Example of a two-second time window of the preprocessed LFP of a PD-induced (i.e., 6-OHDA lesioned) marmoset.
        For a clearer visualisation, signals were bandpass filtered to the [8,50] Hz interval, only for this panel.
        (b) Top two rows show the mean power density spectra (PSD) over all segments for the healthy (blue) and PD (red) marmosets (data for each individual marmoset is included as supplementary material). For thalamic regions (i.e., VL and VPL) and STN, only 6-OHDA lesioned hemispheres are represented, since these regions were not recorded in the healthy marmoset. PSDs were normalised by the maximum PSD value for each time window.
        The bottom row shows the ratio (R) between PD and healthy PSD for each frequency (see Equation \ref{eq:r}).
        To improve visualisation, $\epsilon$ is set to the median of the healthy spectrum.
        a.u.: arbitrary units.}
    \label{fig:spectral_animal-all}
\end{figure*}

Based on two-seconds-long segments, computed according to the data preprocessing steps described in Section~\ref{ssec:data-preprocessing}, (see Figure~\ref{fig:spectral_animal-all}a for a sample), the PSDs of LFPs from healthy and PD marmosets were computed (see Figure~\ref{fig:spectral_animal-all}b for the average spectrum). In all regions of the PD subjects, an increased PSD magnitude from 5 Hz to 25 Hz was observed, which is in accordance with the reported electrophysiological signatures of PD \cite{Tinkhauser2017BetaMedication}. 

From the estimated LFP power spectra from PD marmosets, the target LFP power spectra values for the computational marmoset model were computed as in Equation~\ref{eq:y_r}. The results, presented in Table~\ref{table:targets}, were fed to the DE fitness function (Equation~\ref{eq:fitness}).

\begin{table*}[!ht]
\centering
\caption{Target LFP power spectra values for the computational marmoset model, calculated from data from marmoset monkeys in the PD condition (Equation~\ref{eq:y_r}).}
\label{table:targets}
\vspace{6pt}
\begin{tabular}{@{}llllllll@{}}
\toprule
$y_{StrD1}$ & $y_{StrD2}$ & $y_{TH}$ & $y_{GPi}$ & $y_{GPe}$ & $y_{CtxRS}$ & $y_{CtxFSI}$ & $y_{STN}$ \\ \midrule
%0.2872 & 0.2872 & 0.2155 & 0.2630 & 0.2616 & 0.2624 & 0.2624 & 0.2090 \\ \bottomrule
0.44 & 0.44 & 0.38 & 0.46 & 0.42 & 0.39 & 0.39 & 0.37 \\ \bottomrule
\end{tabular}
\end{table*}

\subsection{Evolutionary algorithm successfully found high fitness genotypes}
\label{ssec:evolution-process}

{After running the DE $N_M=150$ times, the resulting set of high fitness individuals $\mathcal{G}$ (i.e., the highest fitness individual in the population at the end of each of the 60 generations at each evolutionary run)} was analysed. The fitness values of all individuals were recorded at all generations of each evolutionary run.

\begin{figure*}[!ht]
    \centering
    \includegraphics[scale=0.50]{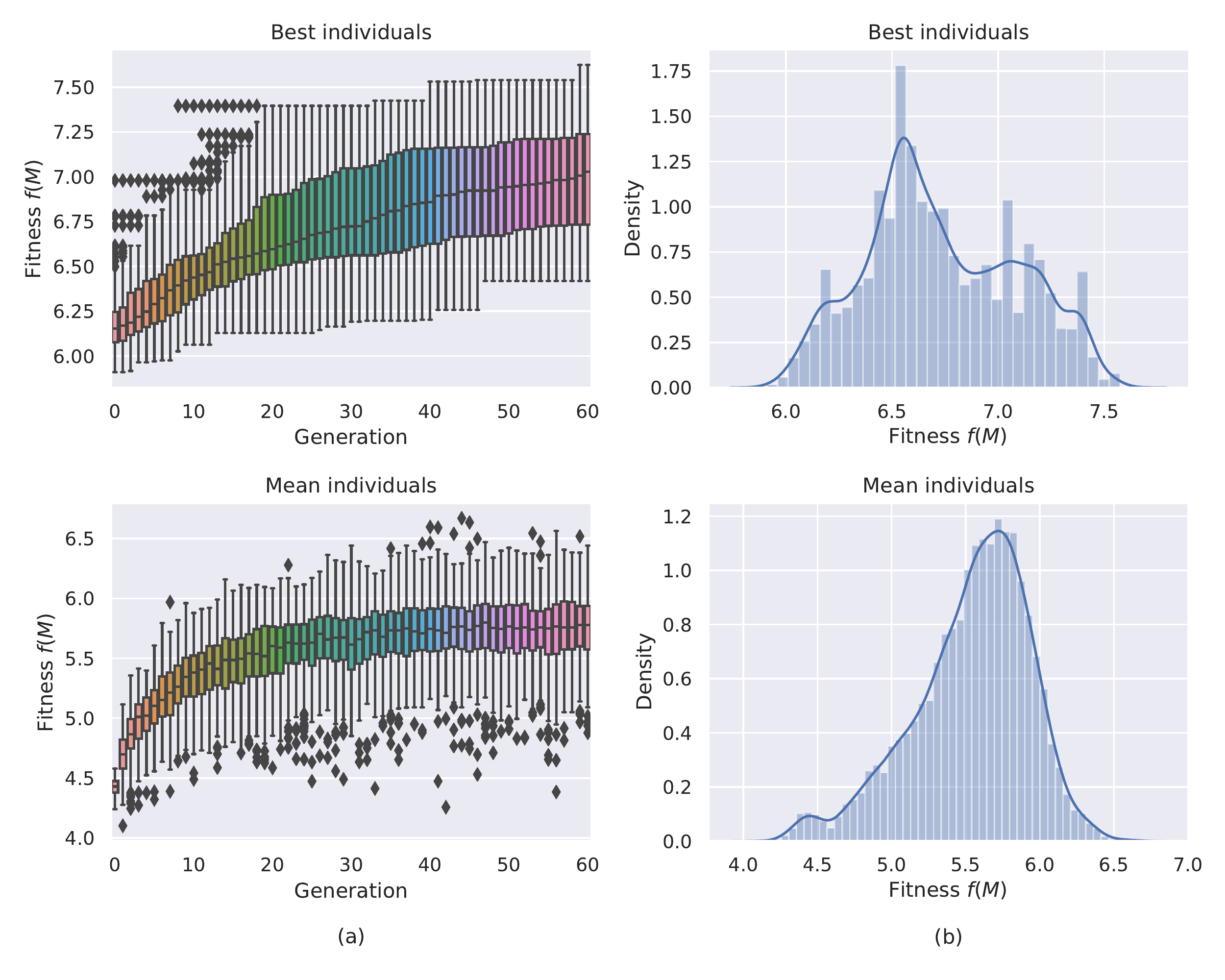}
    \caption{Fitness values  $f(M)$ (Equation~\ref{eq:fitness}) per generation of the evolutionary algorithm (box plots summarising the $k=150$ runs at each generation). The genotypes (i.e., parameter sets for the free parameters elicited in Table~\ref{table:params}) were meant to maximise $f$, which, by definition, would be upper bounded at $8.0$. The upper row refers to the highest fitness individuals at each evolutionary run, and the lower row refers to the mean fitness values of all individuals. (a) Box plots of the best (upper row) and mean (lower row) fitness values at each generation. Outliers were represented by black diamonds. (b) Probability distribution of the best (upper row) and mean (lower row) fitness.}
    \label{fig:fitness}
\end{figure*}

Figure~\ref{fig:fitness} reports the best and mean individual fitness across generations, and the distribution of those values at the end of the evolutionary runs.
{Concretely, the best individual in a given generation is the set of parameters that led to the highest fitness value according to Equation \ref{eq:fitness}. The mean individual fitness across generations refers to the average fitness of all individuals achieved at each generation.}

%Figure~\ref{fig:fitness} report the best and mean individual fitness across generations, and the distribution of those values at the end of the evolutionary runs. {Concretely, the best individual in a given generation is the set of parameters that led to the highest fitness value according to Equation \ref{eq:fitness} set with the target values present in Table~\ref{table:targets}, obtained from actual data from animal models. The mean individual fitness across generations refers to the average fitness value achieved in each generation.}

Regarding the best individual fitness curve, results show that, at every evolutionary run, the initial population contained at least one individual with fitness value close to 6, and that value improved by approximately 1 at the end of evolution (the maximum fitness value possible is $8.0$, see  Equation~\ref{eq:fitness}). Considering the whole population, the initial average fitness was low (approximately 4.5), reaching a plateau close to 5.75 as evolution progressed. The mean fitness across individuals and the best individual's fitness have marginal improvement after generation 40, thus the DE was stopped at $N_{gen}=60$ generations.

\begin{figure*}[ht]
    \centering
    \includegraphics[scale=0.52]{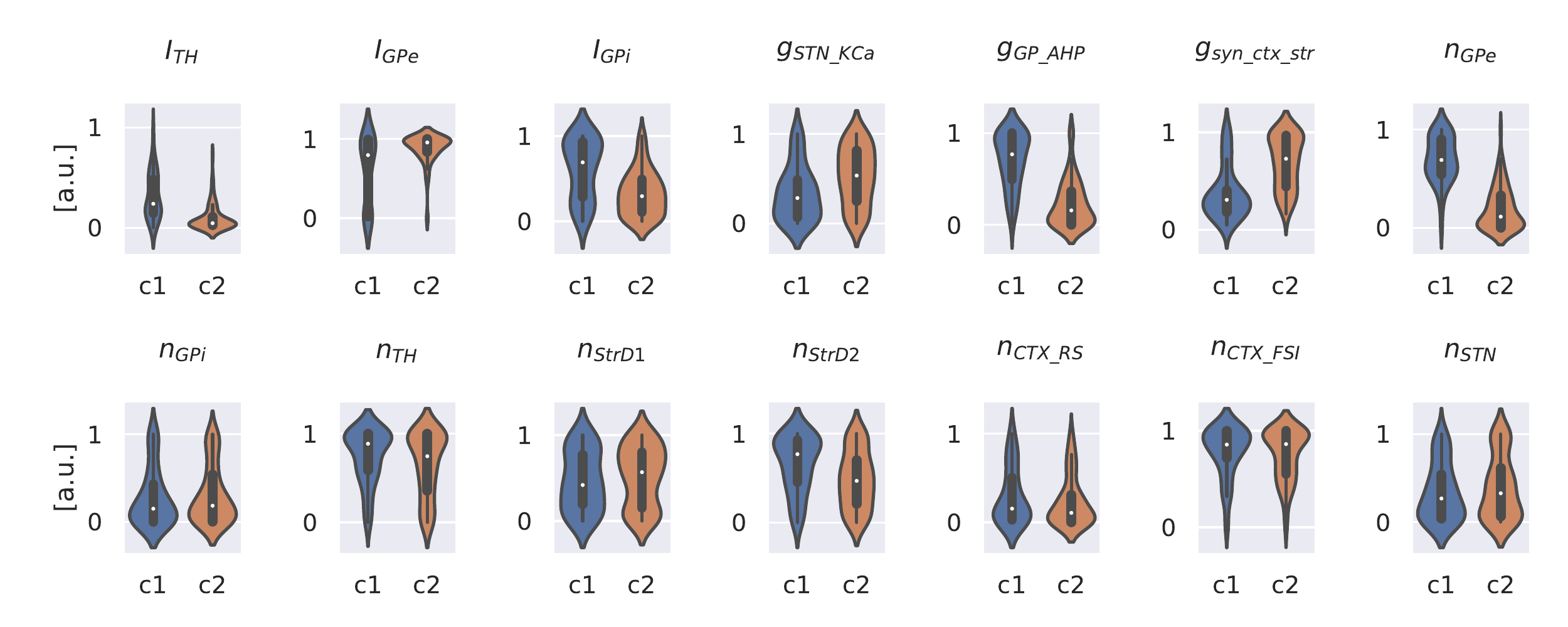}
    \caption{Violin plots showing the distribution of each free parameter (see Table~\ref{table:params}) across the best individuals found at each run of the evolutionary algorithm employed for optimising this set of parameters (i.e., genotype). Although scales vary across parameters (see Table~\ref{table:params}), all parameters were linearly scaled (i.e., normalised) to the interval $[0,1]$ at evolution time. For example, for parameters 7-14 (i.e., the numbers of neurons), a value of zero corresponds to the lower bound of the parameter interval, that is, 10 neurons. a.u.: arbitrary units.}
    \label{fig:params}
\end{figure*}

For all $G \in \mathcal{G}$, we looked at the distribution of parameter values for clusters $C_1$ and $C_2$, represented in Figure~\ref{fig:params}.
Both clusters present similar distributions for most of the parameters, either with small variance (e.g., the numbers of neurons at the cortex populations) or more uniform distributions with high variance (e.g., the number of neurons at the striatum). Other parameters, such as $I_\text{TH}$ and $I_\text{GPe}$, had a clear mean peak and reduced variance in the distribution for $C_2$, but a large variance for $C_1$.

\subsection{High fitness genotypes form two clusters}
\label{ssec:results-evolution-process}

{A set of 150 high fitness individuals was generated by repeatedly running the evolutionary algorithm with different seeds. It is possible that high fitness individuals do not have a unique parameter distribution, and diverse parameter settings could lead to high fitness values. To investigate this issue, we performed a clustering analysis based on the evolved individuals.}

%{Although a set of 150 high fitness individuals was generated by repeatedly running the evolutionary algorithm with different seeds, not necessarily all those individuals could be considered good approximations of the animal models, since different parameter settings could lead to high fitness values. This was the reason for us to include the clustering approach based on these parameter settings: to perform a spectral analysis and, based on it, select the subset of individuals that best approximates the healthy and parkinsonian conditions based on theoretical knowledge \cite{Tinkhauser2017BetaMedication}.}

\begin{figure*}[ht]
    \centering
    \includegraphics[scale=0.85]{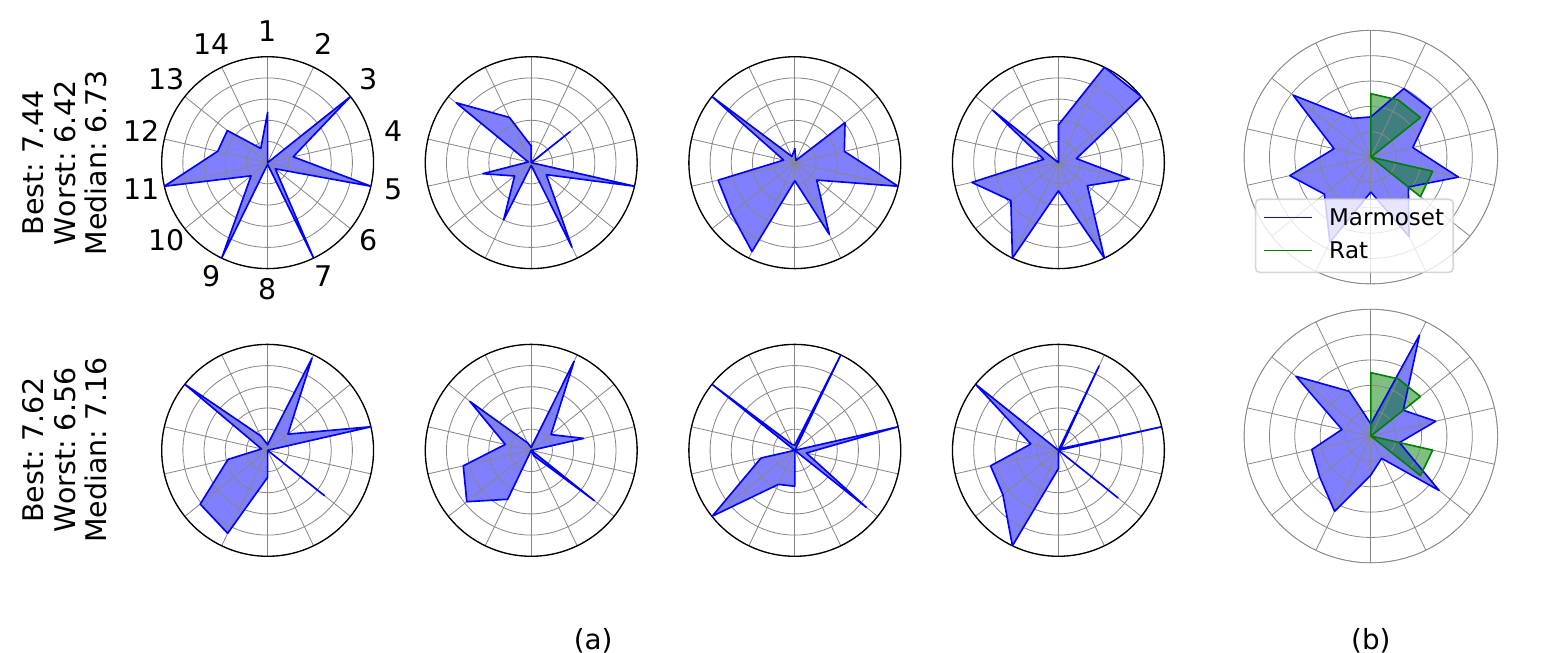}
    \caption{Radar representations of the genotypes (i.e., sets of parameters, see Table~\ref{table:params}) from individuals at each cluster obtained by applying the K-means algorithm, applying these parameters as features for the clustering technique.
    Although scales vary across parameters (see Table~\ref{table:params}), all parameters were linearly scaled (i.e., normalised) to the interval $[0,1]$ at evolution time. For example, for parameters 7-14 (i.e., the numbers of neurons), a value of zero corresponds to the lower bound of the parameter interval, that is, 10 neurons.
    The first row represents cluster $C_1$ and the second row cluster $C_2$. (a) Four individuals with the highest silhouettes with respect to each cluster. Data at the left refers to the fitness $f$ computed as in  Equation~\ref{eq:fitness}. (b) Comparison between the parameters of the rat model and the mean values from each cluster. As in Figure~\ref{fig:params}, parameter values were scaled to the ranges shown in Table~\ref{table:params}, except parameter 4 ($\bar{g_\text{STN\_KCa}}$) of the rat model, whose original value is 1.0 $mS/cm^2$.}
    \label{fig:genotypes}
\end{figure*}

Following the methods from Section~\ref{ssec:clustering-analysis}, the  K-means algorithm was employed to determine $p=2$ clusters. Figure~\ref{fig:genotypes} provides a radar plot representation of genotypes learnt for each cluster, and the correspondence between the mean value of each parameter and those of the rat computational model by Kumaravelu \textit{et al.}~\cite{Kumaravelu2016ADisease}.

Figure~\ref{fig:genotypes}a shows 4 representative genotypes $c_p[1,\dots,4]$, chosen based on the highest silhouettes with respect to each cluster.
For comparison, the parameters from the rat model \cite{Kumaravelu2016ADisease} are superposed with the mean values between all individuals from both clusters in Figure~\ref{fig:genotypes}b. This representation highlights substantial differences between clusters. For instance, the $I_{GPe}$ is at its maximum value in $C_2$, while it shows a much lower value for $C_1$. On the other hand, the number of neurons at the GPe is higher in $C_1$ than in $C_2$.

\subsection{Healthy and PD spectral signatures from computational model resembles those from marmoset monkeys}
\label{ssec:spectral-analysis}

\begin{figure*}[!ht]
    \centering
    \includegraphics[scale=0.75]{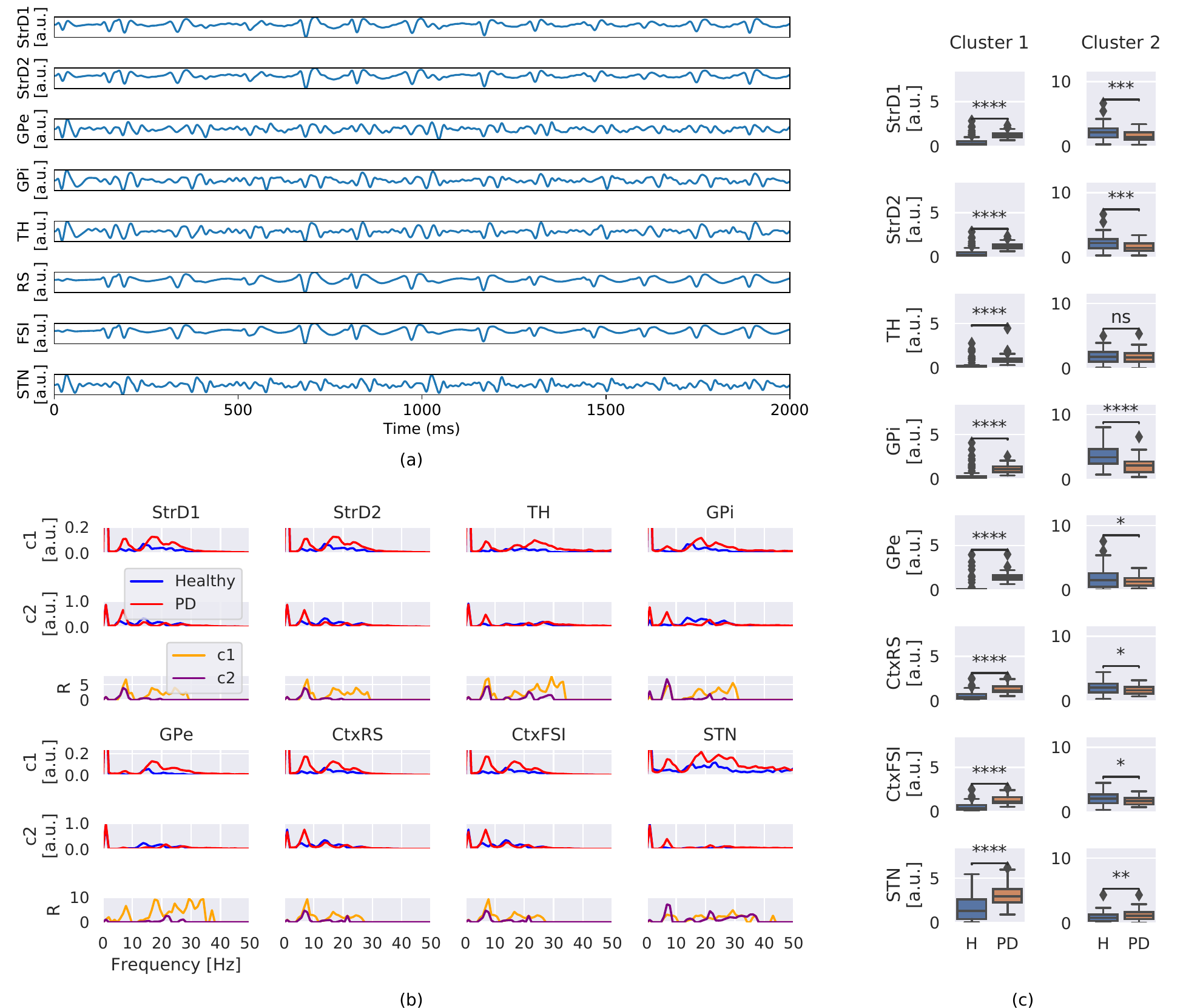}
    \caption{
    Extracellular activity simulated by the computational models resulting from the parameters optimised (i.e., genotypes computed with the evolutionary algorithm), modelled as local field potentials (LFP) at the centre of the regions involved in the BG-T-C circuit. The clusters $C_1$ and $C_2$ were computed by applying the K-means technique directly to the genotypes, hence were not influenced by the simulated neurophysiological activity.
    (a) Example of simulated LFPs for the highest silhouette evolved individual from cluster $C_1$, PD condition.
    For a clearer visualisation, signals were bandpass filtered to the [8,50] Hz interval, only for this panel.
    Compare with Figure~\ref{fig:spectral_animal-all}a.
    (b) Mean PSD for healthy (blue) and PD (red) conditions from the 50 models with the highest silhouette of each cluster, normalised by maximum PSD value for each time window, followed by the ratio $R$ between PD and healthy PSD for each frequency (see Equation \ref{eq:r}).
    %Cluster 1 behaves as expected, with peak activities at the beta range of the PD subjects.
    To improve visualisation, $\epsilon$ is set to the percentile $80$ of the healthy spectrum. 
    (c) Box plot regarding the beta band (13-30 Hz) of the LFP from the 50 models with the highest silhouette of clusters 1 (left) and 2 (right). 
    %The larger beta power in PD models is more prominent in cluster $C_1$.
    Outliers were represented by black diamonds. Unpaired t-tests were applied to evaluate statistical significance against the null hypothesis that H and PD values are drawn from the same underlying distribution (p-value notation:
    $p > 0.05 \rightarrow \text{ns}$;
    $p \in [0.01, 0.05] \rightarrow \text{*}$; % 1.00e-02 < p <= 5.00e-02
    $p \in [0.01, 0.001] \rightarrow \text{**}$; % 1.00e-03 < p <= 1.00e-02
    $p \in [0.001, 0.0001] \rightarrow \text{***}$; % 1.00e-04 < p <= 1.00e-03
    $p < 0.0001 \rightarrow \text{****}$). a.u.: arbitrary units.}
    \label{fig:spectral_CM-all}
\end{figure*}

Regarding the spectral analyses of simulated sessions of the computational model, we employed the same procedure for normalisation as we did for the spectra of the animal model (see Figure~\ref{fig:spectral_animal-all}); that is, we normalised each data segment by the maximum value. 
The sample signals of Figure~\ref{fig:spectral_CM-all}a, shown as an example, were bandpass-filtered to the same range as in Figure~\ref{fig:spectral_animal-all}a to the interval [8-50] Hz.
The mean spectral power and the ratio $R$ are shown for the healthy and PD conditions for each cluster in Figure~\ref{fig:spectral_CM-all}b (see Equation~\ref{eq:r}).

%For an in-depth comparison between the LFP power spectra from the marmoset computational model and real marmoset monkeys, we first constructed PSD box-plots from the animal model data 
In $C_1$, results show higher magnitudes of most frequencies up to 50 Hz for PD models, a fact that is less visible for $C_2$.
The mean PSD ratio from genotypes $G \in C_2$ is close to 1 regardless of frequency range and brain region, whereas genotypes $G \in C_1$ show prominent peaks in beta frequencies. A detailed analysis of box-plots (Figure~\ref{fig:spectral_CM-all}c) confirm the significant differences in the beta band for cluster $C_1$ only.
Considering the spectral densities from the LFP data from animal models (Figure~\ref{fig:spectral_animal-all}b), in which we observe a significant difference in the beta band, results displayed in  Figure~\ref{fig:spectral_CM-all}c confirm that spectral signatures from genotypes in $C_1$ resemble those from marmoset monkeys. Notice that the LFP mean PSDs from the computational model (Figure~\ref{fig:spectral_CM-all}b) has a different shape compared to that from the animal LFP (Figure~\ref{fig:spectral_animal-all}b), but the spectral signature is similar in both healthy and PD conditions and resemble those from marmoset monkeys. This can be explained by the relatively small number of neurons simulated in the computational model \cite{Parasuram2016}.
Therefore, for the forthcoming analyses, only $G \in C_1$ will be considered.

\subsection{Spike activity from healthy models are significantly different from those of PD models}
\label{ssec:mfr-dynamics}

{Regarding spike activity, the marmosets' dataset was not provided with a representative set of spike trains from all regions of the circuit, hence they were not a suitable ground truth for validating the activity from the computational model.
For this reason, the spikes synthesised by the computational model were analysed based on evidence from the literature \cite{Prescott2006AProcessing}}.

\begin{figure*}[!ht]
    \centering
    \includegraphics[scale=0.72]{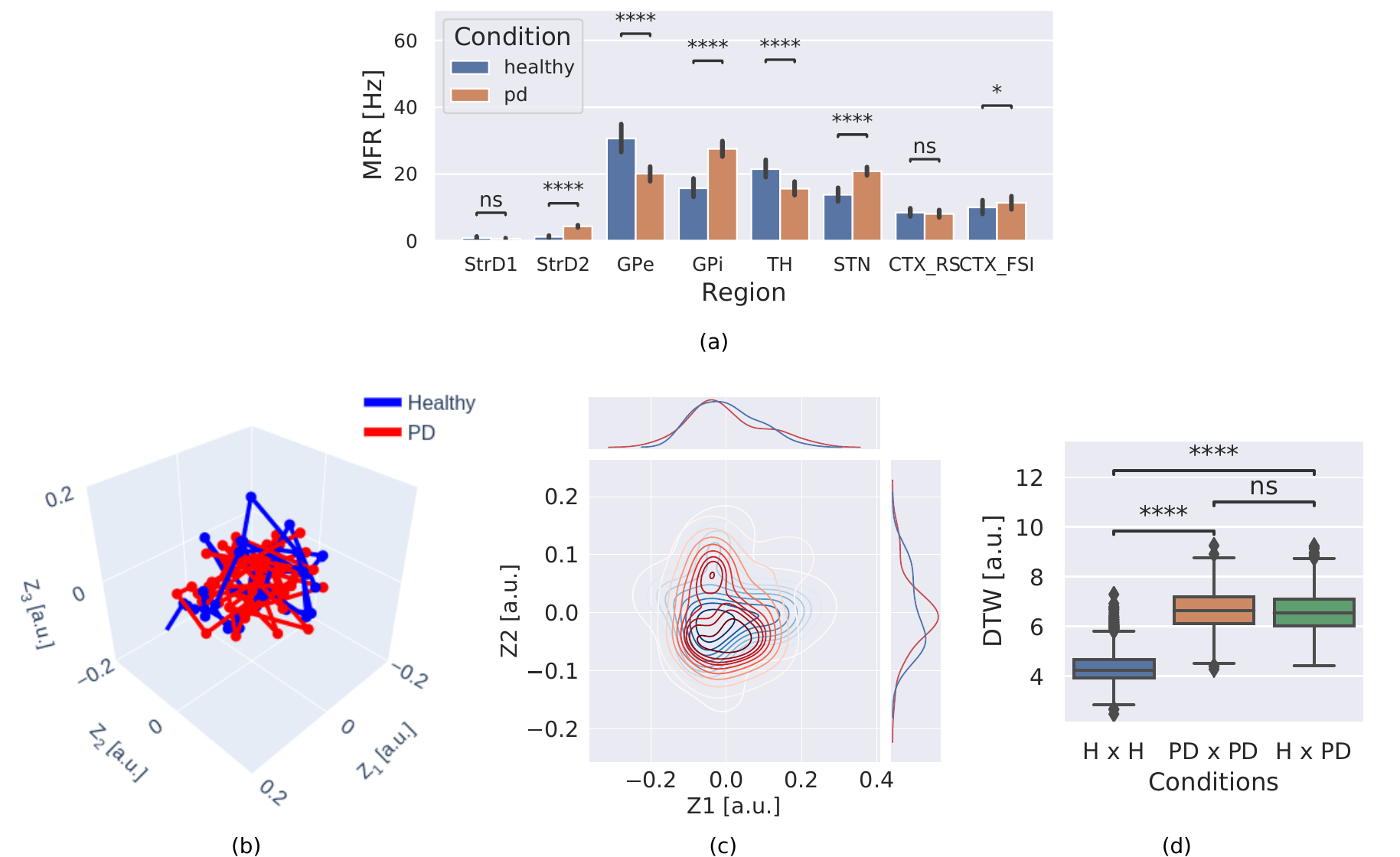}
    \caption{Firing rates and dynamics regarding the spike activity simulated with the computational models derived by the parameter sets from cluster $C_1$. Simulations were ran for $t_{sim}=2000ms$. (a) Mean firing rates for each region in cluster (means and standard deviations). (b) Projection of three principal components of the most representative individual (i.e., highest silhouette) of cluster $C_1$, where $Z_1$, $Z_2$ and $Z_3$ are the principal components with the highest variance. (c) Representation of those components using contour lines. (d) Box plot of the DTW between the dynamics of one simulation of all genotypes belonging to cluster $C_1$. All simulations were performed with the same seed for the generation of random numbers. Higher DTW values mean that the pairs of trajectories being compared are less similar to each other.
    Unpaired t-tests were applied to evaluate statistical significance in (a), against the null hypothesis that H and PD MFR values at each region are drawn from the same underlying distribution, and in (d), against the null hypothesis that a given pair of DTW vectors is drawn from the same distribution as each of the others (p-value notation:
    $p > 0.05 \rightarrow \text{ns}$;
    $p \in [0.01, 0.05] \rightarrow \text{*}$; % 1.00e-02 < p <= 5.00e-02
    %$p \in [0.01, 0.001] \rightarrow \text{**}$; % 1.00e-03 < p <= 1.00e-02
    $p \in [0.001, 0.0001] \rightarrow \text{***}$; % 1.00e-04 < p <= 1.00e-03
    $p < 0.0001 \rightarrow \text{****}$). a.u.: arbitrary units.}
    \label{fig:dynamics-analysis}
\end{figure*}

First, we assessed the differences in mean firing rates (MFR) between the healthy and PD conditions for the marmoset-based computational models in cluster $C_1$. Figure~\ref{fig:dynamics-analysis}a shows the simulated MFR in each brain region for $t_{sim}=2000$ ms, considering the 50 models in $C_1$ with the highest silhouette with respect to the cluster. Results indicate a counter-intuitive relationship between the MFR and the LFP power spectra observed in Figure~\ref{fig:spectral_CM-all}c. Consider, for instance, the GPe and GPi. Both regions show a higher beta-band LFP magnitude in PD condition, but while GPi MFR in PD condition is higher than that from healthy condition, GPe MFR is the opposite.

From Figure~\ref{fig:dynamics-analysis}b and Figure~\ref{fig:dynamics-analysis}c, we observe that neuronal trajectories are intertwined, with no clear difference in the reduced-dimension state space. This is justified by the relatively mild, though statistically significant, differences in MFR (Figure~\ref{fig:dynamics-analysis}a). As described in Section~\ref{ssec:spikeLFP-analysis}, neuronal trajectories were compared with DTW in three scenarios: healthy vs healthy models (HxH), PD vs PD models (PDxPD), and healthy vs PD models (HxPD). As $len(C_1)=53$, the number of pairs from which the DTW was computed was $len(DTW_{C_1}) = \binom{53}{2} = 1378$ for each scenario. The results from this analysis are shown in Figure~\ref{fig:dynamics-analysis}d, in which the scalar outputs of the DTW algorithm are considered for all possible pairs within groups, for the  HxH and PDxPD comparisons, or between groups, for the HxPD comparisons. Since two trajectories generated by the same individual were not compared on any of the analyses, we have computed statistical significance using unpaired tests, differently from the remaining analyses in the paper.

Trajectories from the HxH scenario were statistically more similar than trajectories from the other conditions. Thus, the intertwined trajectories observed in PCA (Figure~\ref{fig:dynamics-analysis}b and Figure~\ref{fig:dynamics-analysis}c) in fact relate to significant differences between healthy and PD neuronal dynamics. Interestingly, PDxPD trajectories differ more than those from HxH, which can be interpreted as a less homogeneous, regarding neuronal dynamics, genotype to phenotype mapping.

\subsection{Healthy and PD spike coherences from the computational model resemble that from marmoset monkeys}
\label{ssec:coherence}

\begin{figure*}[!ht]
    \centering
    \includegraphics[scale=0.3]{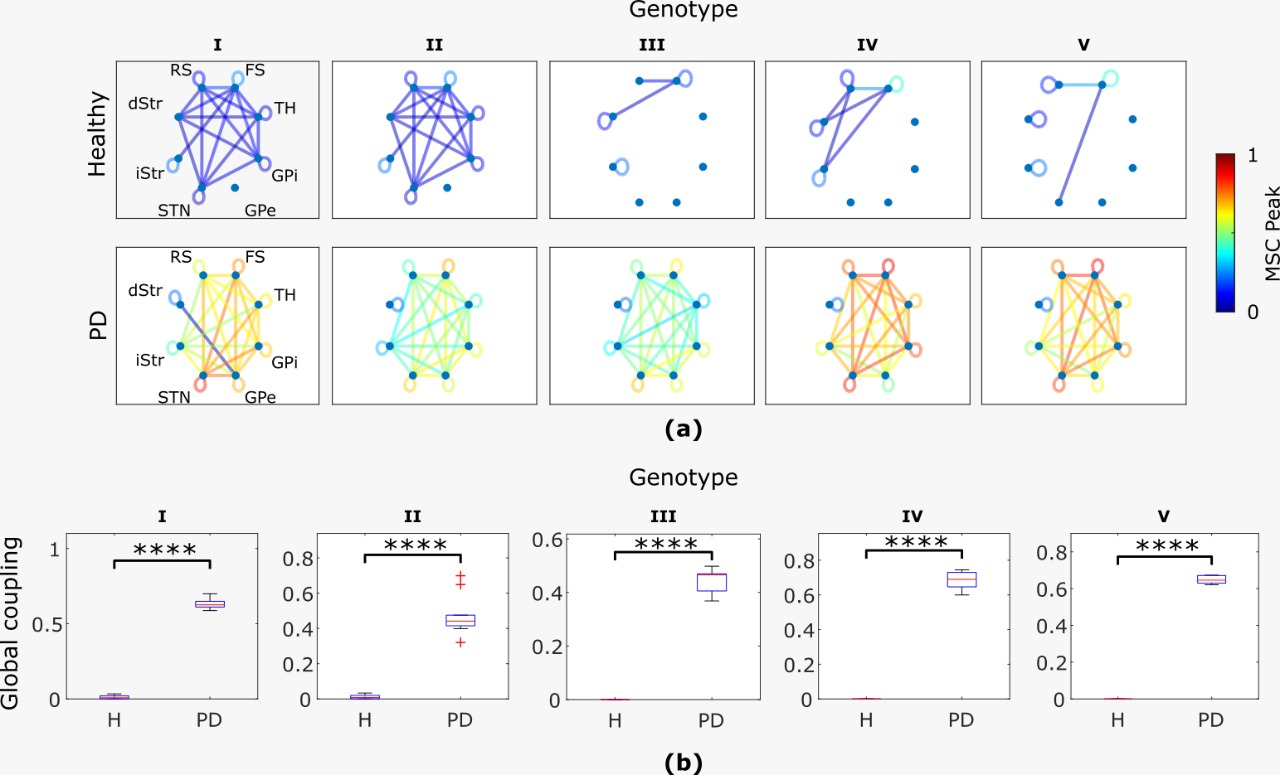}
    \caption{Coherence analyses computed for the spike activity of the five parameter sets, optimised through the evolutionary algorithm, with the highest silhouettes with respect to cluster $C_1$. These parameter sets were used to construct the healthy and PD computational models that were considered in the coherence analyses. (a) Peak magnitude-squared coherence (MSC) in the 13-30 Hz band within and between each simulated brain region for the top five genotypes with highest silhouette from cluster $C_1$. Only connections whose peak MSC values are above significance level are shown. (b) Global coupling metric (median value of the MSC matrix) between brain regions for healthy and PD conditions (see Section~\ref{ssec:model-validation} and Equation \ref{eq:coh}). (p--value notation:
    %$p > 0.05 \rightarrow \text{ns}$;
    %$p \in [0.01, 0.05] \rightarrow \text{*}$; % 1.00e-02 < p <= 5.00e-02
    %$p \in [0.01, 0.001] \rightarrow \text{**}$; % 1.00e-03 < p <= 1.00e-02
    %$p \in [0.001, 0.0001] \rightarrow \text{***}$; % 1.00e-04 < p <= 1.00e-03
    $p < 0.0001 \rightarrow \text{****}$)}
    \label{fig:coherence}
\end{figure*}

To conclude our model validation, we selected the top five genotypes with highest silhouette from cluster $C_1$ and calculated the magnitude-squared coherence (MSC) within and between each simulated brain region (Section~\ref{ssec:coherence}) for healthy and PD conditions. Results revealed that computational models ran in the healthy condition provided a lower peak MSC in the 13-30 Hz band when compared to that from the PD condition (Figure~\ref{fig:coherence}a), with two important observations: genotype I has higher peak MSC when compared to the other 4 genotypes in the healthy condition, and genotypes II and III have a lower widespread peak MSC when in the PD condition compared to that from other genotypes in the same condition. Statistical analysis confirmed the significant differences in all five genotypes when comparing the global coupling metric (see Section~\ref{ssec:model-validation} and Equation \ref{eq:coh}) between healthy and PD conditions (Figure~\ref{fig:coherence}b); that is, PD models present a higher widespread coherence in the 13-30 Hz band than that observed in healthy models.

\section{Discussion}
\label{sec:discussion}

%hyperactivity indirect pathway -> striatum is silent -> against what is expected in PD 

%calabrese direct indirect pathways nature neuroscience
%10.1038/nn.3743

Marmoset monkeys are prominent in neuroscience research~\cite{Cyranoski2009,Kishi2014,Mitchell2015,Miller2016}. Although there are anatomical and physiological differences between BG-T-C circuit in rodents and primates, neurophysiological data from rodents are far more available than from primates. Considering the similarities on the overall structure of the BG-T-C circuit among all vertebrates~\cite{Koprich2017AnimalDevelopment}, we assumed that the rat model presented by Kumaravelu \textit{et al.}~\cite{Kumaravelu2016ADisease} was a suitable starting point to build a computational model of those structures in primates. The core hypothesis was that, by keeping the same brain regions and connectivity patterns of the rat model and modifying a set of parameters, the computational model could reproduce neural dynamics of healthy and PD marmoset conditions.

The dataset used in this work included simultaneous LFP recordings from regions of the BG-T-C network and power spectra (PSD) analysis revealed significantly higher 13 to 30 Hz LFP PSD magnitudes for PD marmosets in all regions. This result might be interpreted cautiously, given that one healthy marmoset is being compared to three PD marmosets. Also, results refer to a broad range of frequencies, hence different interval choices may influence the analysis. Nonetheless, one would expect a widespread significant increase in LFP power centred in (but not limited to) the beta band in PD affected brains~\cite{Santana2014SpinalDisease,Tinkhauser2017BetaMedication}. 

Regarding the MFR results from the computational model (Figure~\ref{fig:dynamics-analysis}a), there are significant differences between the healthy and PD conditions. Single-neuron firing rates vary considerably depending on animal species, whether the animal is fully awaken, engaged in behavioural tasks, or anaesthetised~\cite{Hardman2002,VanAlbada2009,McGregor2019CircuitDisease}). Data from human subjects, even though scarce, are in line with animal results \cite{Du2018}. Moreover, there is a great neuronal diversity within the BG-T-C network, both in terms of neuronal physiology and connectivity, which have been shown to have a non-trivial relationship with field potentials~\cite{Sharott2009,Bolam2000,Holgado2010,Benhamou2012}. Our model takes into account this diversity, which was shown in Figures \ref{fig:dynamics-analysis}b and \ref{fig:dynamics-analysis}c; nevertheless, the reported MFR are in agreement with the literature: comparing PD with healthy conditions, a higher MFR in GPi, STN, and Str, and a lower MFR in GPe, TH, and CTX. 

The data-driven modelling strategy adopted in this paper is consolidated in computational neuroscience literature~\cite{Nowke2018}, but often leads to multiple models fitting a particular data set~\cite{Bahuguna2017}. Therefore, model optimisation should be followed by a model selection phase. We clustered high fitness solutions with respect to evolved parameters and obtained two clusters, and found two clear sets of parameters that reproduce the increased beta-band oscillations observed in PD marmosets~\cite{Santana2014SpinalDisease}. However, when perturbing the model to shift from PD to healthy dynamics, only one of the clusters fitted the marmoset data. Notably, we evolved solutions based on LFP data but computational model firing rates resemble those reported in previous works~\cite{VanAlbada2009,Li2015,Du2018}. Nevertheless, as data becomes available, future works should explore different fitness functions based on single-neuron activities or other features of LFP.
Lastly, in this context, our simulated neurons are formed by a single cylindrical compartment, thus future works should consider using neurons with more complex compartments and connections, possibly including multiple dendritic branches and active ionic channels. This would lead to more realistic simulated LFP signals \cite{Parasuram2016}, but at the expense of heavier computing resources.

One of the great challenges in neuroscience is to link the activity of large neural populations to motor and cognitive behaviours. One strategy is to study the intrinsic high-dimensional dynamics of neural populations from its low-dimensional dynamics given by time-varying trajectories~\cite{Cunningham2014,Sussillo2014}, thus emphasising circuit over single-neuron function. For example, Humphries \textit{et al.}~\cite{Humphries2018} showed that neural low-dimensional dynamics given by PCA of neuronal activity can explain \textit{Aplysia} rhythmic movement control and propose that only the low-dimensional dynamics are consistent within and between nervous systems. Also, the shape and amplitude of neural trajectories can explain different behavioural outcomes \cite{Gamez2019}. Combining PCA and DTW, we found that neural trajectories from high-fitness models are more similar in healthy conditions than in PD conditions. This is in line with results from Russo \textit{et al.}~\cite{Russo2019}, who demonstrated, using computer simulations, later confirmed by data from the supplementary motor area in monkeys, that low trajectory divergence is essential in neural circuits involved in action control. PCA is a simple, established method for dimensionality reduction, but other computational tools tailored to neuronal data, such as Gaussian-Process Factor Analysis (GPFA)~\cite{Yu2009} and jPCA~\cite{Churchland2012}, should be considered in further analyses. Another possible approach is to use more advanced machine learning methods to identify PD-related features from neural data, as demonstrated by Ranieri \textit{et al.}~\cite{Ranieri2020}, who employed a deep learning framework to unveil PD features from marmoset data.

Finally, as part of our model validation, we assessed functional coupling within and between simulated brain regions by means of coherence between spike trains. In contrast to structural coupling, characterised by physical neuronal connections, functional connectivity is an emergent phenomenon commonly linked to synchronisation in neural rhythms at diverse spatiotemporal scales and is the basis of neural communication and cognitive processing~\cite{Singer1999,Buzsaki2010,Fries2015,Lakatos2019}. Several neural disorders, including PD, present a disruption in functional connectivity \cite{Uhlhaas2010,Mathalon2015,Halje2019}. 
In particular, Santana \textit{et al.} \cite{Santana2014SpinalDisease} showed that 6-OHDA marmoset models of PD have a widespread coherence peak in the beta band when compared to healthy individuals. Our computational model is in line with this result, which is relevant not only as further evidence of its biological plausibility, but also because one of the established therapies to alleviate PD motor symptoms is the use of deep brain stimulation (DBS) \cite{Poewe2017}. 
%In particular, Santana \textit{et al.}~\cite{Santana2014SpinalDisease} showed that 6-OHDA marmoset models of PD have a widespread coherence peak in the 8–15 Hz range in relation to the 30–40 Hz band when compared to healthy individuals. Our computational model is in line with this result, which is relevant not only as further evidence of its biological plausibility, but also because one of the established therapies to alleviate PD motor symptoms is the use of deep brain stimulation (DBS)~\cite{Poewe2017}.
Thus, we believe that the work presented here can be used to test hypotheses that employ DBS. For instance, Romano \textit{et al.}~\cite{Romano2020} performed a comprehensive analysis of frequency-dependent effects of DBS on the same model that we used here, tuned for rodent data \cite{Kumaravelu2016ADisease}, and found that neural oscillatory modulations were similar to those observed in electrical brain and spinal cord stimulation of primates~\cite{Wang2018,Santana2014SpinalDisease}.

{Certain simplifications inherent to our approach should also be mentioned, as they may serve as inspiration for improvements in future research. In our work, LFP generation followed the method described in Parasuram \textit{et al.}~\cite{Parasuram2016}, and implemented in NetPyNE, which does not consider the influence of sinks. Despite being a simplification, the method has been able to reproduce features of real LFP data, and is computationally feasible. In this approach, LFP waveforms are directly related to transmembrane ionic currents from each neuronal source, which in turn relate to neuronal firing rates, and electrode position. As we have assigned coordinates to the simulated electrodes corresponding to the centre of each simulated region, we can assume that simulated LFP dynamics are due to altered spiking activity in multiple neuronal sources from different brain regions.}

{Likewise Kumaravelu \textit{et al.}~\cite{Kumaravelu2016ADisease} and previous seminal BG-T-C modelling works such as Humphries \textit{et al.}~\cite{Humphries2006}, and van Albada and Robinson~\cite{VanAlbada2009}, we did not model any structural synaptic plasticity mechanisms. Our synapses were modelled as bi-exponential and alpha synapses, including transmission delays. Nevertheless, as model dynamics unfold, functional plasticity mechanisms may take place in the sense that the closed-loop, recursive network architecture could lead to single neurons and brain regions whose electrical activity are sensitive to past network states. In fact, the depletion of dopamine, one of the hallmarks of PD, affects structural and functional plasticity. Our model considers the loss of dopaminergic neurons (see Section~\ref{ssec:CM}, for a complete description), thus we believe that the model is suited for the investigation of functional plasticity phenomena. This analysis is beyond the scope of our work, but the reader can relate the change in oscillatory neural dynamics we described to different functional states. For instance, Humphries \textit{et al.}~\cite{Humphries2006} show that action selection in the BG is closely linked to oscillatory activity.}

There are several directions for future work. Based on the study of Wang \textit{et al.}~\cite{wang2016subthalamic}, a phase amplitude coupling analysis in the STN in our computational model may shed light on the different aspects of LFP oscillations observed in healthy and PD conditions, as well as the mechanisms underlying these oscillations. In another perspective, most PD computational models do not consider brain-body-environment interactions. Embodied cognitive science studies have provided solid evidence that neural activity is shaped by such interactions~\cite{Engel2001,Pfeifer2006,Badcock2019,Musall2019}. In PD and other neural disorders, body-environment interactions influence motor control~\cite{Snijders2010,Santos2017}, but its impact on neural dynamics remains unclear. Moreover, the BG-T-C neuronal network is clearly related to action selection and decision making~\cite{Humphries2006,Mink2018,Suryanarayana2019}. Therefore, we believe that associating our marmoset-based computational model with a physical robot may offer an alternative approach to elucidate the mechanisms underlying brain-body-environment interactions in PD~\cite{Gurney2004,Prescott2006,Krichmar2018,Pronin2021NeuroroboticReview, Pimentel2021}. A possible approach would be to employ this computational model in a sensorimotor loop based on visual inputs from video cameras and motor outputs to actuators such as the robot's motors. In this scenario, computer vision algorithms would transform the images into stimuli for the computational model, so that the resulting currents and action potentials would be used to generate perturbations that would govern the behaviours of the actuators. The resulting framework could become a new tool for studying the underlying mechanisms of PD and the effects of different interventions regarding the simulated circuit.

%In future works, we plan to use this model in a neurorobotics context, in which sensory inputs and motor responses can be used to highlight functional plasticity mechanisms differences between healthy and PD states.

\section{Conclusions}
\label{sec:conclusion}

Computational models are invaluable tools for advancing our knowledge of the neural dynamics of our brain, either under healthy conditions or with neurological disorders. Even though the physiopathology underlying PD shares similarities across vertebrate species, there are important, species-specific differences in the anatomy and neural dynamics of the BG-T-C circuit. For example, the number of neurons in the GPe is considerably increased in primate models, when compared to rodent models. Hence, the design of a primate computational model of PD is of paramount importance. 
In this work, we created the first computational model of the dynamics of BG-T-C motor circuit based on data from Marmoset monkeys both in healthy and parkinsonian conditions. Our data-driven approach used simultaneous, multisite electrophysiological recordings from healthy and 6-OHDA+AMPT marmoset models of PD.
We are aware that there are simplifications in our computational model; nevertheless, results show that LFP power spectral densities at frequencies of interest, firing frequency dynamics, and spike coherence resemble those from healthy and PD marmosets.

Electrophysiological datasets from animal models often do not include comprehensive biophysical data such as single-neuron membrane conductances and neuronal cell densities. These parameters were central for building a biophysical computational model. Thus, to address this gap, we used an optimisation algorithm (differential evolution) to search the multidimensional model parameter space for solutions that could reproduce features of the animal LFP recordings.
Our model was based on a well known rat model of PD \cite{Kumaravelu2016ADisease}. The main aspects of novelty in our modelling approach are: 1) we use a marmoset monkey BG-T-C electrophysiological database; 2) we added LFP simulations to the model, in addition to spike dynamics; and 3) we developed a DE-based optimisation to search for unknown parameters. With this framework, we were able to reproduce several of the previously reported PD electrophysiological biomarkers observed and recorded from marmoset monkeys.

{Our computational model presents beta-band LFP power spectra differences between the healthy and the PD conditions, which Wang \textit{et al.}~\cite{wang2016subthalamic} also found in human patients with dystonia.
This is in line with a body of literature that shows that beta-band LFP modulations are not a PD-specific biomarker (see Poewe \textit{et al.}~\cite{Poewe2017} and references therein).
Although our model is focused on PD, the electrophysiological features we use are known to be related to other neural disorders and thus should not be considered as exclusive to PD.}

\section*{Acknowledgement}

This work is part of the Neuro4PD project, granted by Royal Society and Newton Fund (NAF\textbackslash R2\textbackslash180773), and S\~ao Paulo Research Foundation (FAPESP), grants 2017/02377-5 and 2018/25902-0. Moioli and Araujo acknowledge the support from the National Institute of Science and Technology, program Brain Machine Interface (INCT INCEMAQ) of the National Council for Scientific and Technological Development(CNPq/MCTI), the Rio Grande do Norte Research Foundation (FAPERN), the Coordination for the Improvement of Higher Education Personnel (CAPES), the Brazilian Innovation Agency (FINEP), and the Ministry of Education (MEC). Romano was the recipient of a master's scholarship from FAPESP, grant 2018/11075-5. Elias is funded by a CNPq Research Productivity Grant (312442/2017-3). This research was carried out using the computational resources from the Center for Mathematical Sciences Applied to Industry (CeMEAI) funded by FAPESP, grant 2013/07375-0.
Additional resources were provided by the Robotics Lab within the Edinburgh Centre for Robotics, and by the Nvidia Grants program.

\bibliographystyle{plain}
\bibliography{references}

\end{document}